\documentclass[twocolumn,prb,superscriptaddress]{revtex4-1}
\usepackage{graphicx,bm,amsmath,amssymb,color,soul,xcolor,tikz,siunitx,braket,cases,cleveref,algpseu
docode,soul}

\let\hide\iffalse

\renewcommand{\cite}[1]{{\color{magenta}[#1]}}

\def\bp{{\bf p}}
\def\bk{{\bf k}}
\def\br{{\bf r}}

\def\bq{{\bf q}}
\def\a{{\alpha}}
\def\b{{\beta}}
\def\bu{{\bf u}}
\def\be{{\bf e}}
\def\bR{{\bf R}}

\def\btau{{\bm \tau}}
\def\D{\partial}
\def\d{\delta}
\def\w{\omega}

\def\bG{{\bf G}}
\def\bT{{\bf T}}

\def\<{\langle}
\def\>{\rangle}
\def\k{\kappa}
\def\ve{\varepsilon}
\def\e{\epsilon}

\def\F{Fr\"ohlich}

\def\bqin{{\bf q}_\parallel}
\def\bQin{{\bf Q}_\parallel}
\def\bQ{{\bf Q}}
\def\Qin{Q_\parallel}
\def\brin{{\bf r}_\parallel}

\def\bGin{{\bf G}_\parallel}
\def\qin{{q}_\parallel}

\begin{document}

\title{Unified \textit{ab initio} description of Fr\"ohlich electron-phonon \\[2pt] 
interactions in two-dimensional and three-dimensional materials}

\author{Weng Hong Sio} 
\affiliation{Institute of Applied Physics and Materials Engineering, University of Macau, 
Macao SAR 999078, P. R. China}
\affiliation{Oden Institute for Computational Engineering and Sciences, The University of 
Texas at Austin, Austin, Texas 78712, USA}

\author{Feliciano Giustino}
\email{fgiustino@oden.utexas.edu}
\affiliation{Oden Institute for Computational Engineering and Sciences, The University of 
Texas at Austin, Austin, Texas 78712, USA}
\affiliation{Department of Physics, The University of Texas at Austin, Austin, Texas 
78712, USA}

\date{\today}

\begin{abstract} 
\textit{Ab initio} calculations of electron-phonon interactions including the polar Fr\"ohlich 
coupling have advanced considerably in recent years. The Fr\"ohlich electron-phonon matrix element 
is by now well understood in the case of bulk three-dimensional (3D) materials. In the case of 
two-dimensional (2D) materials, the standard procedure to include Fr\"ohlich coupling is to employ 
Coulomb truncation, so as to eliminate artificial interactions between periodic images of the 
2D layer.  While these techniques are well established, the transition of the 
Fr\"ohlich coupling from three to two dimensions has not been investigated. Furthermore, it remains 
unclear what error one makes when describing 2D systems using the standard bulk 
formalism in a periodic supercell geometry. Here, we generalize previous work on the \textit{ab 
initio} Fr\"ohlich electron-phonon matrix element in bulk materials by investigating the 
electrostatic potential of atomic dipoles in a periodic supercell consisting of a 2D material and a continuum dielectric slab. We obtain a unified expression for the matrix element, which reduces to the existing formulas for 3D and 2D systems when 
the interlayer separation tends to zero or infinity, respectively. This new expression enables an 
accurate description of the Fr\"ohlich matrix element in 2D systems without resorting 
to Coulomb truncation.  We validate our approach by direct \textit{ab initio} density-functional 
perturbation theory calculations for monolayer BN and MoS$_2$, and we provide a simple expression 
for the 2D Fr\"ohlich matrix element that can be used in model Hamiltonian approaches. 
The formalism outlined in this work may find applications in calculations of polarons, quasiparticle 
renormalization, transport coefficients, and superconductivity, in 2D and quasi-2D materials.
\end{abstract}

\maketitle

\section{Introduction}

The electron-phonon interaction (EPI) plays an important role in many materials 
properties,\citep{Giustino2017} including the carrier mobility of 
semiconductors,\citep{Balandin2012,Ishiwata2013} phonon-assisted optical 
processes,\citep{Noffsinger2012,Zacharias2015,Novko2019} vibrational spectroscopy,\citep{Eliel2018, 
Miller2019,Cong2020} polaron physics,\citep{Devreese2009,Verdi2017,Sio2019prl, Husanu2020} and 
superconducting pairing.\citep{Oliveira1988,Margine2016} During the past decade, calculations of 
EPIs have become more accessible, and much work has been performed on the role of phonons in the 
optical and transport properties of semiconductors and other functional 
materials.\citep{Villegas2016,Fan2018,Kang2019, Hinsche2017,Ponce2019prb,Ponce2020review, 
Ponce2020} Given the significant interest in two-dimensional (2D) materials and their 
applications,\citep{Geim2011, Andersen2015,Mounet2018} \textit{ab initio} calculations of EPIs in 
2D systems are also becoming increasingly 
popular.\citep{Sohier2016,Kaasbjerg2012,Sohier2018,Deng2021,Li2019} 

The key element of \textit{ab initio} calculations of EPIs is the electron-phonon matrix element, 
$g_{mn\nu}(\bk,\bq)$, which describes the probability amplitude for an electron to be scattered 
from an initial Bloch state with wavevector $\bk$ and band index $n$ to a final state with 
wavevector $\bk+\bq$ and band index $m$ by a phonon of wavevector $\bq$ and branch index $\nu$. In 
the majority of known three-dimensional (3D) semiconductors and insulators, this matrix element 
diverges as $1/|\bq|$ for small $\bq$, as a result of the long-range nature of the electric field 
generated by fluctuating atomic dipoles. This singular behavior is referred to as the Fr\"ohlich 
electron-phonon coupling,\citep{Frohlich1950} and occurs whenever the atoms in a crystal exhibit 
nonvanishing Born effective charges.\citep{Verdi2015, Sjakste2015,Vogl1976}

Calculations of EPIs in bulk 3D systems including the Fr\"ohlich coupling are well established by 
now,\citep{Verdi2015,Sjakste2015} and are routinely performed in conjunction with Wannier-Fourier 
interpolation.\citep{Giustino2007,Marzari2012} In the case of 2D materials, several proposals have 
been put forward for dealing with the Fr\"ohlich EPI, including parametrized model matrix 
elements,\citep{Kaasbjerg2012, Ma2020} calculations using the formalism for 3D 
systems,\citep{Li2019} and fully \textit{ab initio} approaches employing Coulomb 
truncation.\citep{Sohier2018,Deng2021} All these approaches focus on the case of a 
monolayer system embedded in a vacuum buffer in periodic supercell calculations.  More complex 
configurations, including van der Waals heterostructures, semiconductor/insulator interfaces, and 
moir\'e bilayers,\citep{Bistritzer2011, Cao2018,Chen2019moire,Mcgilly2020} are still beyond the 
reach of current methods. Furthermore, the connection between current approaches for 2D systems and 
the previous theory for 3D systems remains unclear. In order to enable the study of EPIs in a 
broader class of materials and their interfaces, it is desirable to develop a single unified 
framework for describing Fr\"ohlich EPIs in 3D and 2D systems on the same footing.

At a more fundamental level, there is also the question on how to connect \textit{ab initio} 
calculations of EPIs in 2D systems with model Hamiltonian approaches.  Earlier work considered the 
so-called ``strict 2D limit'' of the Fr\"ohlich matrix element, whereby electrons are assumed to be 
confined in a sheet of vanishing thickness.\citep{Sarma1985, Peeters1985} This limit was 
successfully employed to investigate polarons and quasiparticle renormalization in 2D 
systems,\citep{Sak1972,Mason1986,Jalabert1989,Sarma1990,Hahn2018} but in this model the matrix 
element diverges as $|\bq|^{-1/2}$ at small $\bq$.  This behavior contrasts with the fact that, in 
realistic systems with small but finite thickness, the long-wavelength limit of the Fr\"ohlich 
matrix elements is finite.\citep{Mori1989} This inconsistency poses a challenge when attempting to 
relate the results derived from earlier models and even recent diagrammatic Monte Carlo 
studies\citep{Hahn2018} to atomic-scale \textit{ab initio} calculations of EPIs.

Here we address these difficulties by developing a unified Fr\"ohlich EPI matrix element which 
seamlessly describes 3D and 2D systems within a periodic supercell geometry. The present approach 
is a generalization of the approach of Ref.~\onlinecite{Verdi2015} for bulk 3D systems. While 
Ref.~\onlinecite{Verdi2015} derived the Fr\"ohlich matrix element by examining the electrostatics 
of a dipole in a bulk crystal, here we examine the potential generated by a dipole within a 2D slab 
embedded in a uniform dielectric medium (such as vacuum, for example).

We validate this approach by comparing our analytical expressions with explicit density-functional 
perturbation theory (DFPT) calculations for monolayer BN and MoS$_2$, and we show that our method 
reproduces the expressions of Ref.~\onlinecite{Verdi2015,Sjakste2015,Vogl1976} in the 3D limit, as 
well as the expression of Ref.~\onlinecite{Sohier2016} in the limit of large interlayer separation 
between the periodic images of the 2D layer.

The manuscript is organized as follows. In Sec.~\ref{sec:frohlich} we discuss the formalism to 
describe Fr\"ohlich EPIs in bulk and 2D materials. In particular, in 
Sec.~\ref{sec:frohlichcoupling_bulk} we review the basics of Fr\"ohlich coupling in bulk 3D 
materials from the point of view of \textit{ab initio} calculations, and the connection between the 
\textit{ab initio} formalism and the analytical model originally derived by Fr\"ohlich. In 
Sec.~\ref{sec:frohlichcoupling_strict2d} we briefly summarize existing approaches to Fr\"ohlich 
coupling in 2D systems, the underlying assumptions, and their limitations. In Sec.~\ref{sec:theory} 
we derive a new expression for the \textit{ab initio} Fr\"ohlich matrix element in 2D and quasi-2D 
systems.  In Sec.~\ref{sec.limits} we show how our expression recovers the 3D matrix element of 
Refs.~\onlinecite{Verdi2015,Sjakste2015, Vogl1976} (Sec.~\ref{sec.bulklimit}) and the 2D 
matrix element of Ref.~\onlinecite{Sohier2016} (Sec.~\ref{sec.sohier}) in the respective 
limits. In the same section we also derive an expression for the special case of 
atomically thin single-layer crystals in vacuum, where all the atoms lie in the same plane 
(Sec.~\ref{sec:model}). In Sec.~\ref{sec.simplemodel} we rewrite our main results from 
Sec.~\ref{sec.limits} in a form that depends on only macroscopic quantities and that is 
particularly suitable for use in model Hamiltonian approaches. Section~\ref{sec:results} reports 
applications of this methodology to monolayer BN and MoS$_2$. In particular, in
Sec.~\ref{sec:computational_details} we provide details on the computational setup and the 
optimized materials parameters.  In Sec.~\ref{sec:validation} we calculate the \textit{ab 
initio} Fr\"ohlich matrix elements in monolayer BN and monolayer MoS$_2$, and validate our 
method by direct comparison with DFPT calculations.  In Sec.~\ref{sec:model2dfrohlich} we examine 
the dependence of the polar coupling on the size of the vacuum gap using our closed-form 
expressions and \textit{ab initio} materials parameters for BN. In Sec.~\ref{sec:conclusion} we 
summarize our findings and offer our conclusions.  

\section{The Fr\"ohlich electron-phonon matrix element in 3D and 2D systems: Earlier work} 
\label{sec:frohlich}

\subsection{Fr\"ohlich coupling in bulk 3D solids} \label{sec:frohlichcoupling_bulk}

In this section we first recall the expression for the 3D Fr\"ohlich matrix element as derived in 
Ref.~\onlinecite{Verdi2015}. The same expression was obtained in Refs.~\onlinecite{Sjakste2015, 
Vogl1976} following a different route. Then we clarify the connection between the \textit{ab 
initio} \F\ matrix element and the classic result by \F.

The EPI matrix element can be written \citep{Giustino2017} as $g_{mn\nu}(\bk,\bq) = 
\braket{\psi_{m\bk+\bq}| \Delta_{\bq\nu}V|\psi_{n\bk}}$, where $\psi_{n\bk}$ and $\psi_{m\bk+\bq}$ 
are typically Kohn-Sham wavefunctions, and $\Delta_{\bq\nu}V$ is the linear variation of the 
Kohn-Sham potential associated with a phonon of frequency $\w_{\bq\nu}$. $\Delta_{\bq\nu}V$ can be 
calculated using DFPT\citep{Baroni2001} or the frozen phonon method.\citep{Yin1982}

Reference~\onlinecite{Vogl1976} showed that, in crystals with a finite gap between occupied and 
unoccupied states, the EPI matrix element can be expanded in a Laurent series near $\bq=0$.  This 
series may contain terms that scale as $\mathcal{O}(q^{-1})$, $\mathcal{O}(q^0)$, 
$\mathcal{O}(q^1)$, and so on, where $q=|\bq|$. The $\mathcal{O}(q^{-1})$ term corresponds to an 
electric dipole potential, the $\mathcal{O}(q^0)$ term corresponds to a quadrupole, the term 
$\mathcal{O}(q)$ is for an octopole, and so on. In materials with nonzero Born effective charges, 
such as for example polar semiconductors and oxides, the dipole term is nonzero and dominates in 
the limit $q\rightarrow 0$.  Since the dipole and quadrupole terms are nonanalytic near $q=0$, 
these terms must be treated separately when performing Wannier interpolation of the EPI matrix 
elements. The \textit{ab initio} procedure to deal with the dipole term was developed in 
Refs.~\onlinecite{Verdi2015,Sjakste2015}, and the corresponding procedure for dealing with the 
quadrupole term was reported recently in~Refs.~\onlinecite{Brunin2020prl,Brunin2020prb,Park2020, Jhalani2020}. In all cases one 
writes the matrix element as:
\begin{equation}
g_{mn\nu}(\bk,\bq) = g_{mn\nu}^{\mathcal{S}}(\bk,\bq) + g_{mn\nu}^{\mathcal{L}}(\bk,\bq),
\end{equation}
where the superscripts stand for short- and long-range, respectively. 
$g_{mn\nu}^{\mathcal{L}}(\bk,\bq)$ contains all nonanalyticities, and is designed to capture the 
exact limit of $g_{mn\nu}(\bk,\bq)$ for $\bq\rightarrow 0$. The form of 
$g_{mn\nu}^{\mathcal{L}}(\bk,\bq)$ away from $\bq=0$ is inconsequential, as long as it is a smooth 
function of the phonon wavevector.

In the following we focus on the $\mathcal{O}(q^{-1})$ component of 
$g_{mn\nu}^{\mathcal{L}}(\bk,\bq)$, which is commonly known as the Fr\"ohlich interaction. The 
extension of the present formalism to deal with quadrupoles is possible, at least in principle, but 
this would require a separate investigation. For notational simplicity, below we drop the 
superscript in $g_{mn\nu}^{\mathcal{L}}(\bk,\bq)$, and we use $g_{mn\nu}(\bk,\bq)$ to indicate the 
Fr\"ohlich component of the matrix element.

To obtain the Fr\"ohlich matrix element, Ref.~\onlinecite{Verdi2015} proceeded in two steps: (1) 
evaluate the electrostatic potential generated by a point dipole $\bp$ in an anisotropic medium 
characterized by the high-frequency relative dielectric permittivity tensor $\bm{\epsilon}^\infty$, and (2) associate one such dipole to every atom $\k$ in the unit cell with lattice vector $\bR$, 
undergoing the displacement:
\begin{equation}\label{eq.disp}
\Delta \btau_{\kappa \bR}^{(\bq\nu)} = (\hbar/2 M_\kappa \omega_{\bq\nu})^\frac{1}{2}e^{i\bq \cdot 
\bR} \bf{e}_{\kappa \nu}(\bq).
\end{equation}
Here, $M_\k$ is the atomic mass and $\bf{e}_{\kappa \nu}(\bq)$ is the vibrational eigenvector 
normalized in the unit cell.  This atom carries the (dimensionless) Born effective charge tensor 
$\bm{Z}^*_\kappa$.  The resulting matrix element is \citep{Verdi2015,Giustino2017err}
\begin{eqnarray} \label{eq:verdi}
&&g_{mn\nu}(\bk,\bq) = i \frac{4\pi}{\Omega} \frac{e^2}{4\pi \ve_0} \sum_\kappa \bigg( 
\frac{\hbar}{2M_\kappa \omega_{\bq\nu}} \bigg)^{\frac{1}{2}} \sum_{\bG \neq -\bq}  \nonumber \\&& 
\frac{ (\bq+\bG) \cdot \bm{Z}^*_{\kappa} \cdot \be_{\kappa \nu}(\bq) }{(\bq+\bG)\cdot 
\bm{\epsilon}^\infty \cdot (\bq+\bG)} \braket{\psi_{m\bk+\bq} |  e^{i(\bq+\bG)\cdot 
(\br-\btau_{\kappa})} | \psi_{n\bk}},\quad
\end{eqnarray}
where $\Omega$ is the volume of the primitive unit cell, and $\btau_\kappa$ is the equilibrium 
position of this atom. $\bG$ denotes the reciprocal lattice vectors, and the braket indicates the 
integral over the Born-von K\'arman (BvK) supercell.

The classic matrix element by \F\ can be obtained from Eq.~\eqref{eq:verdi} by considering the 
following approximations. (1) We consider $\bq$ in the first Brillouin zone, so that the only 
singularity is at $\bq=0$ and the summation over $\bG$ can be ignored. (2) For all quantities that 
vary smoothly with $\bq$, we retain only the corresponding $\bq=0$ limit.  (3) The band structure 
is described using the electron gas model,  so that  $\psi_{n\bk}(\br) = (N\Omega)^{-1/2} e^{i\bk 
\cdot \br}$.  (4) Phonons are described using the Einstein model, therefore there are two 
transverse optical (TO) branches with $\omega_{\bq\nu} = \w_{\rm TO}$ and one LO branch with 
$\omega_{\bq\nu} = \omega_{\text{LO}}$. (v) The dielectric permittivity tensor is isotropic, 
$\bm{\epsilon}^\infty_{\a\b} = \epsilon^\infty \d_{\a\b}$, with Greek indices denoting Cartesian 
coordinates. Using these approximations in Eq.~\eqref{eq:verdi}, we find
\begin{equation}\label{eq.fr.model1}
|g_{\nu}(\bq)|^2 = \left[\frac{4\pi}{\Omega} \frac{e^2}{4\pi \ve_0}  \frac{1}{\e^\infty}\right]^2 
\frac{\hbar}{2 M_0 \omega_{\rm LO}} \frac{ \left|\bq \cdot {\bm Z}^*_{\nu}\right|^2 }{q^4},
\end{equation}
where we removed the redundant band indices  and we introduced the mode-effective Born charge ${\bm 
Z}^*_{\nu}$ following Ref.~\onlinecite{Gonze1997}, $Z^*_{\nu,\alpha} = \sum_{\k,\b} 
(M_0/M_\k)^{1/2} Z^*_{\k,\a\b} e_{\kappa\b, \nu}(0)$. In these expressions, 
$M_0$ is an arbitrary reference mass that is introduced to keep $Z^*_{\nu\,a}$ dimensionless. The 
matrix element in Eq.~\eqref{eq.fr.model1} depends on the angle between $\bq$ and ${\bm 
Z}^*_{\nu}$. By performing a spherical average over this angle (taking into account the volume 
element in three dimensions), and summing over the LO/TO manifold, we obtain a single effective matrix element:
\begin{equation}\label{eq.fr.model2}
|g(q)|^2 = \left[\frac{4\pi}{\Omega} \frac{e^2}{4\pi \ve_0}  \frac{1}{\e^\infty}\right]^2 
\frac{\hbar}{2 M_0 \omega_{\rm LO}} \frac{1}{q^2} \sum_\nu \left|{\bm Z}^*_{\nu}\right|^2.
\end{equation}
The sum on the right hand side is related to the static and high-frequency dielectric 
permittivities by:\citep{Gonze1997} 
 \begin{equation}\label{eq.fr.model3}
 \e^0 = \e^\infty +  \frac{e^2}{4\pi \ve_0}\frac{4\pi}{\Omega} \frac{\sum_\nu |{\bm Z}^*_\nu|^2}{ 
M_0  \w_{\rm TO}^2}.
 \end{equation}
Using this expression and the Lyddane-Sachs-Teller relation, $ \e^0/\e^\infty =\w_{\rm 
LO}^2/\w_{\rm TO}^2$, Eq.~\eqref{eq.fr.model2} can be rewritten in the standard 
form:\citep{Frohlich1950}
\begin{equation}\label{eq.fr.model5}
 |g(q)| = \alpha^{1/2}_{\text{FR}}  \hbar\w_{\rm LO} \frac{q_{\rm FR}}{q},
\end{equation}
where the dimensionless \F\ coupling strength $\alpha_{\text{FR}}$ is defined 
as:\citep{Devreese2009}
 \begin{equation} \label{eq:couplingstrength3d}
 \alpha_{\text{FR}} = \frac{e^2}{4\pi \ve_0} \frac{1}{\hbar} \sqrt{\frac{m^*}{2\hbar 
\omega_{\text{LO}}}} \bigg( \frac{1}{\epsilon_\infty} - \frac{1}{\epsilon_0} \bigg),
 \end{equation}
$m^*$ is the band effective mass, and $q_{\rm FR}$ is a characteristic wavevector given by $q_{\rm 
FR}^2 = 4\pi \Omega^{-1} (\hbar/2 m^* \w_{\rm LO} )^{1/2}$. Equations~\eqref{eq:verdi} and 
\eqref{eq.fr.model5} show that, in three dimensions, the \F\ interaction diverges as $1/q$, as is well known.

\subsection{\F\ coupling in 2D systems} \label{sec:frohlichcoupling_strict2d}

In early studies of polar electron-phonon coupling interactions in 2D systems, the \F\ matrix 
element was derived either within the strict 2D limit,\citep{Peeters1985} or by 
considering electrons confined within an infinite square well potential along the direction 
perpendicular to the slab ($z$ direction in the following).\citep{Sarma1985}

The square-well approximations are arrived at by considering that, in calculations of physical 
properties, the 3D electron-phonon matrix element is modulated by the electron density along 
the $z$ direction:\citep{Ercelebi1987,Titantah2001,Ponce2020review,Hahn2018,Sarma1985}
\begin{equation} \label{eq:2dstep}
|g^{\text{2D}}(\bqin)|^2 = \frac{c}{2\pi} \int_{-\infty}^\infty dq_z \,\,  F(q_z)
|g^{\text{3D}}(\bqin,q_z)|^2,
\end{equation}
where $\bqin$ and $q_z$ are the components of the phonon momentum $\bq$ parallel and perpendicular 
to the slab, respectively, $c$ is the slab thickness, and $F(q_z)$ is the Fourier component of the 
electron density profile along $z$, $F(z)$:
\begin{equation}
F(q_z) = \int_{-\infty}^\infty dz \,\,  F(z) e^{-iq_z z}. 
\end{equation}
When the electron is strictly confined in a 2D sheet of zero thickness, the profile 
becomes a Dirac delta function, $F(z) = \delta(z)$, and one obtains the 2D \F\ matrix elements 
in the strict 2D limit.\citep{Hahn2018,Sarma1985,Kaasbjerg2012} After integrating out the third 
dimension in Eq.~(\ref{eq:2dstep}), the matrix element is written as:\citep{Sarma1985,Hahn2018}
\begin{equation}
 |g^{\rm 2D}(\qin)| = \alpha^{1/2}_{\text{FR}} \hbar\w_{\rm LO} \frac{q_{\rm FR} (c / 2)^{1/2}}{\qin^{1/2}},
\end{equation}
A similar result can alternatively be derived starting from the Coulomb potential in two 
dimensions.\citep{Peeters1985} At variance with the standard 3D \F\ matrix element, which scales as 
$q^{-1}$, the matrix elements in the strict 2D approximation scales as $\qin^{-1/2}$.  This 
singular behavior is currently understood to be an artifact of the model, which is inconsistent 
with experiments.\citep{Rucker1992} To overcome this limitation, Ref.~\onlinecite{Kaasbjerg2012} 
employed a Gaussian profile of width $\sigma$ to described the electron density along the $z$ 
direction. The resulting matrix element in the long-wavelength limits reads
\begin{equation}\label{eq.Kaas}
|g^{\text{2D}}(\bq_{||})| = g \ \text{erfc}\left(\qin\sigma/2\right),
\end{equation}
where the constants $g$ and $\sigma$ are determined by fitting this expression to \textit{ab 
initio} data. This model has successfully been employed to investigate the transport properties of 
transition-metal dichalcogenide monolayers.\citep{Kaasbjerg2012} One potential limitation of this 
approach is that the relation between the electron-phonon coupling strength $g$ and materials 
parameters such as dielectric constants, Born charges, and vibrational frequencies is not apparent 
as in the standard \textit{ab initio} \F\ matrix element.\citep{Verdi2015,Sjakste2015} Being able 
to trace the coupling back to these properties would be desirable, so as to establish predictive 
analytical models for electron-phonon physics in two dimensions.

To bridge the gap between model studies of electron-phonon interactions in 2D materials and 
first-principle calculations, Ref.~\onlinecite{Sohier2016} developed a refined model which takes 
into account microscopic features such as Born charges, phonon frequencies, and the dielectric 
permittivity of the 2D slab. In this model one assumes that the atomic displacements generate a 
uniform macroscopic polarization density within the slab.  For brevity we quote the expression 
obtained by Ref.~\onlinecite{Sohier2016} for a 2D slab surrounded by vacuum (the general expression 
can be found in Ref.~\onlinecite{Sohier2016}):
\begin{eqnarray} \label{eq:sohier}
|g^{\text{2D}}(\qin)| &=& \frac{2\pi}{A}\frac{e^2}{4\pi\ve_0} \left[ {\sum}_{\kappa}\frac{\hbar}{2 
M_\k \w_{\rm LO}}Z^{*,2}_{\kappa,\parallel} \right]^{1/2} \,\frac{1}{\epsilon_\infty} \frac{2}{\qin 
d}\nonumber \\ &\times& \left[ 1 + \frac{\epsilon_{\infty}^{-1}}{\qin d}  \frac{e^{\qin d} - 1}{ 
1-(1+\epsilon_{\infty}^{-1})(1+e^{\qin d})/2}  \right].
\end{eqnarray}
In this expression, $d$ and $A$ are the slab thickness and unit-cell area, respectively, $\epsilon_{\infty}$ 
is the (isotropic) high-frequency permittivity, $\w_{\rm LO}$ is the frequency of the 
longitudingal-optical (LO) mode, and $M_\k$ and $Z^{*}_{\kappa,\parallel}$ are the atomic masses 
and in-plane Born charges, respectively. As for Eq.~\eqref{eq.Kaas}, the coupling given by 
Eq.~\eqref{eq:sohier} is not singular for $\qin \rightarrow 0$.

The model leading to Eq.~\eqref{eq:sohier} has successfully been employed in calculations of 
electron-phonon couplings in two dimensions,\citep{Sohier2016,Cheng2018,Sohier2018} and constitutes the 
\textit{de facto} state-of-the-art approach in the field.  There have been at least two 
generalizations of the model of Ref.~\onlinecite{Sohier2016}, which additionally take into account 
the out-of-plane polarization and the effect of exact 2D long-range screening.\citep{Deng2021,Royo2021} However, all these approaches are 
designed to describe a 2D slab between two semi-infinite media.  In some cases it may be desirable 
to model the 2D system as a periodic superlattice rather than an isolated slab, for example when 
studying van der Waals heterostructures or 2D semiconductor/insulator interfaces. In 
the next section we derive such a model for quasi-2D systems, by generalizing the approach 
developed in Ref.~\onlinecite{Verdi2015} for bulk crystals.

\section{Derivation of the Fr\"ohlich matrix element in quasi-2D systems} \label{sec:theory}

\begin{figure}
  \centering
  \includegraphics[width=\columnwidth]{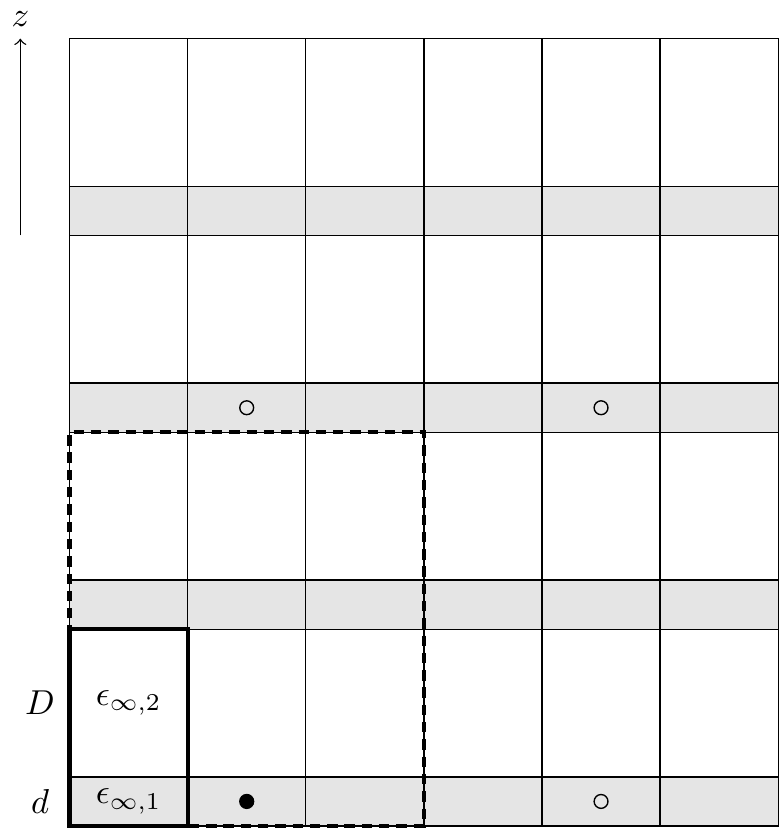}
\caption{Schematic representation of a superlattice consisting of a periodic stack of two layers 
alternating along the $z$ direction. The material under consideration is the layer of thickness $d$ 
and dielectric constant $\epsilon_{\infty,1}$. The other layer, of thickness $D$ and permittivity 
$\epsilon_{\infty,2}$, is a dielectric continuum. The thick solid line denotes the boundary of the primitive 
unit cell; the thick dashed line indicates the boundary of the BvK supercell.  
Equation~\eqref{eq:inhomopot} gives the electrostatic potential generated by a point charge located 
in one of the layers (indicated by the disk $\bullet$). To obtain a solution that is periodic in 
the BvK supercell, we superimpose the potential of all periodic images (indicated by the circle 
$\circ$). From the displacements of these charges we obtain the dipole potential.  Our Fr\"ohlich 
matrix element in Eq.~\eqref{eq:frohlichquasi2d} is then obtained by summing over the dipoles 
associated with every atomic displacement.}
\label{fig:fig1}
\end{figure}

In this section we generalize the reasoning leading to Eq.~\eqref{eq:verdi} to the case of a 
periodic stack where two materials alternate along the $z$ direction, as shown in 
Fig.~\ref{fig:fig1}.  We label these materials as ``primary'' and ``secondary'' layer, 
respectively.  We consider the primary layer to be the 2D slab of interest, and the secondary layer 
to be the embedding medium or subrate. For example the primary layer could be a monolayer of 
MoS$_2$, and the secondary layer could be multilayer BN or vacuum. We describe the secondary layer 
as a homogeneous dielectric medium, \textit{without taking into account the discrete, atomic-scale 
structure of this layer}.

We denote the nominal thickness of the primary and secondary layer as $d$ and $D$, respectively, 
and the unit cell length in the $z$ direction as $c=d+D$.  For convenience we shall say that the 
primary layer extends from $z=-d$ to $z=0$, and the secondary layer occupies the region from $z=0$ 
to $z=D$.  The unit cell is repeated periodically within a Born-von K\'arm\'an (BvK) supercell 
consisting of multiple unit cells in the $xy$ plane and in the $z$ direction, and periodic boundary 
conditions on the BvK cell are applied.

To keep the theory as simple as possible, we assume that the two layers can be described by 
effective isotropic high-frequency (relative) dielectric permittivities $\epsilon_{\infty,1}$ and 
$\epsilon_{\infty,2}$, following the same line or reasoning as in Refs.~\onlinecite{Sohier2016}.  In 
particular, we assume that  the dielectric permittivity is given by:
 \begin{equation}\label{eq:profile}
   \epsilon_{\infty}(z) = \begin{cases} \epsilon_{\infty,1} &  -d<z<0 \\ 
   \epsilon_{\infty,2} & \,\,\,\,0<z<D.  \end{cases}
 \end{equation}
This assumption is legitimate because (1) we are interested only in the long-wavelength limit of 
the electron-phonon matrix element, and (2) the out-of-plane dielectric permittivity can 
effectively be made equal to the in-plane permittivity via an appropriate choice of the thickness 
$d$.\citep{Sohier2016} Evidently there is no sharp boundary between the primary and secondary layer 
in real materials, and the dielectric permittivity evolves smoothly.\citep{Giustino2005} However, 
the notion of a sharp dielectric interface considerably simplifies the equations without affecting the final 
results.

To extend the reasoning of Ref.~\onlinecite{Verdi2015} to the present case, we evaluate the 
electrostatic potential of a point dipole located at the position $\btau$ in the primary layer. To 
this aim, we begin by determining the electrostatic potential of a point charge $e$ located at 
$\btau$, \textit{without} considering its periodic replicas.  Subsequently, we proceed to 
\textit{replicate} this charge in all BvK supercells in order to describe realistic 
first-principles calculations.  Following the notation of Ref.~\onlinecite{Verdi2015} [see 
Eq.~(S1) of that work], the electrostatic potential of a single charge \textit{without} its 
replicas is obtained as the solution of the inhomogeneous Poisson's equation:
\begin{equation} \label{eq:poisson}
\epsilon_{\infty}(z) \nabla^2 \varphi(\br;\tau_z) + \frac{d \epsilon_{\infty}(z)}{dz}  
\frac{\partial \varphi(\br;\tau_z)}{\partial z} = -\frac{e}{\varepsilon_0}\d(\br-\tau_z\bu_z)~,
\end{equation}
where we have set $\tau_x=\tau_y=0$ to start with. $\bu_z$ is the unit vector along $z$, and 
$-d<\tau_z<0$.  To proceed we  perform a Fourier integral for the in-plane coordinates:
\begin{equation} \label{eq:fsv}
\varphi(\br;\tau_z) = \! \!  \int \! d \bQin \,\, \varphi(z,\bQin;\tau_z) e^{i\bQin \cdot \brin},
\end{equation}
where the $\bQin$'s denote in-plane wave vectors and $\brin$ is the position in the $xy$ plane.  
After replacing Eq.~\eqref{eq:fsv} inside Eq.~\eqref{eq:poisson} we obtain the following equation 
for $\varphi(z,\bQin;\tau_z)$:
\begin{equation} \label{eq:1dpoisson}
\epsilon_{\infty}(z) \frac{\partial^2 \varphi}{\partial z^2} + \frac{d\epsilon_{\infty}
(z)}{dz} \frac{\partial \varphi}{\partial z} - \bQin^2 \epsilon_{\infty}(z) \varphi = -\frac{e}{ (2\pi)^2 
\ve_0} \delta(z-\tau_z).
\end{equation}
The exact solution of this equation with the two-step dielectric profile defined in 
Eq.~(\ref{eq:profile}) has been derived by Ref.~\onlinecite{Guseinov1984}, in the context of a 
study of excitons in periodic superlattices. The solution is:
\begin{eqnarray} \label{eq:inhomopot}
&&\varphi(z;\bQin;\tau_z) = \frac{e}{2 (2\pi)^2  \ve_0 \epsilon_{\infty,2} \Qin (\gamma^- - \gamma^+)} 
\times \nonumber\\ && \begin{cases} z >  \tau_z:& \\ \left[(\alpha + \gamma^+ \beta)e^{\Qin \tau_z} 
+ (\beta + \gamma^+\alpha)e^{-\Qin \tau_z}\right]\varphi_-(z,\bQin)~, & \\ z < \tau_z : & \\ 
\left[(\alpha + \gamma^- \beta)e^{\Qin \tau_z} + (\beta + \gamma^-\alpha)e^{-\Qin 
\tau_z}\right]\varphi_+(z,\bQin)~, & \end{cases}\nonumber \\
\end{eqnarray}
where $\Qin=|\bQin|$ and the function $\varphi_{\pm}$ is defined as:
\begin{eqnarray} \label{eq:function}
&&\varphi_{\pm}(z,\bQin) = \\ && \begin{cases} nc-d < z < nc:\\ \quad e^{\pm n\eta}\left[(\alpha + 
\gamma^\pm \beta) e^{\Qin (z-nc)} + (\beta + \gamma^\pm \alpha)e^{-\Qin (z-nc)}\right] \nonumber 
\\[3pt] nc < z < nc+D:\\ \quad e^{\pm n\eta} [e^{\Qin (z-nc)} + \gamma^\pm e^{-\Qin (z-nc)}], 
\end{cases}
\end{eqnarray}
with $n$ being an integer. The quantities $\a$, $\b$, $\gamma^{\pm}$, and $\eta$ appearing in these 
expressions are defined as follows:
\begin{eqnarray}
\alpha &=& (1+\epsilon_{\infty,2}/\epsilon_{\infty,1})/2~, \label{eq.defs1}\\ \beta&=& 
(1-\epsilon_{\infty,2}/\epsilon_{\infty,1})/2~,\\ \gamma^\pm &=& 
-\frac{ \displaystyle e^{\Qin D} - e^{\pm \eta}\left(\alpha e^{-\Qin d} + \beta e^{\Qin d}\right)  
}{ \displaystyle e^{-\Qin D} - e^{\pm \eta}\left(\beta e^{-\Qin d} + \alpha e^{\Qin d}\right)  
}~,\\ \eta &=& \cosh^{-1} \big\{ \cosh[\Qin(D-d)] + 2\alpha^2/(2\alpha - 1) \nonumber \\ &\times& 
\sinh(\Qin D) \sinh(\Qin d)  \big\}.\label{eq.eta} \phantom{\int} \label{eq.defs4}
\end{eqnarray}
For later reference, it is useful to note that the potential $\varphi(z,\bQin;\tau_z)$ transforms 
as follows upon translations of the atomic coordinate by a unit cell vector along the $z$ axis:
\begin{equation}\label{eq.varphi_transl}
\varphi(z,\bQin;\tau_z+R_z) = \varphi(z-R_z,\bQin;\tau_z).
\end{equation}
This property follows immediately from Eq.~\eqref{eq:1dpoisson}. 

In order to determine the complete 3D Fourier transform of the potential, we write:
\begin{equation} \label{eq:fsv2}
\varphi(\br;\tau_z) = \! \!  \int \! d \bQ \,\, \varphi(\bQ;\tau_z) e^{i\bQ \cdot \br},
\end{equation}
and we compare this expression with Eq.~\eqref{eq:fsv} to obtain:
\begin{equation} \label{eq:fourier}
\varphi(\bQ;\tau_z) = \frac{1}{2\pi}\int dz \,\varphi(z,\bQin;\tau_z) e^{-iQ_z z}~.
\end{equation}
As a consequence of Eq.~\eqref{eq.varphi_transl}, the potential $\varphi(\bQ;\tau_z)$ transforms 
like a Bloch function under translation by a unit cell vectors along the $z$ axis:
\begin{equation}\label{eq.varphi_transl2}
\varphi(\bQ;\tau_z+R_z) = e^{iQ_z R_z}\varphi(\bQ;\tau_z).
\end{equation}
At this point we can replace Eqs.~\eqref{eq:inhomopot}-\eqref{eq.eta} inside Eq.~\eqref{eq:fourier} 
and and evaluate the integral. After some algebra we find:
\begin{equation}\label{eq.varphi}
\varphi(\bQ;\tau_z) = \frac{e}{2(2\pi)^3\ve_0 \epsilon_{\infty,2} \Qin} K(\bQ,\tau_z),
\end{equation}
where the kernel function $K(\bQ,\tau_z)$ is a rather involved combination of complex exponentials 
which, besides $\bQ$ and $\tau_z$, depend on  the geometric and dielectric parameters of the stack, 
$d$, $D$, $\epsilon_{\infty,1}$, and $\epsilon_{\infty,2}$.  The complete expression for 
the kernel is provided in Appendix~\ref{app.kernel}; see Eq.~\eqref{eq:kernel}.

The electrostatic potential $\varphi(\br;\tau_z)$ given by Eqs.~\eqref{eq:fsv2} and 
\eqref{eq.varphi} corresponds to a single charge in the dielectric stack. In order to impose BvK 
boundary conditions, we place this charge at $\btau$ and replicate it in every BvK supercell. We 
call the resulting potential $\phi(\br;\btau)$:
\begin{equation}\label{eq.V}
\phi(\br;\btau) = \sum_\bT \varphi(\br-\bT;\btau)~,
\end{equation}
where the $\bT$'s are the the lattice vectors of the BvK supercell.  We note that, to avoid an 
unphysical divergence of the potential, we should add to Eq.~\eqref{eq.V} the potential of a 
neutralizing background.  However, this contribution cancels out when evaluating the potential of 
point dipoles, therefore it can safely be ignored.  Since the potential $\phi(\br;\btau)$ is 
periodic in the supercell, we can expand it in a discrete Fourier series:
\begin{equation}\label{eq.Vf}
\phi(\br;\btau) =\sum_{\bq,\bG\ne -\bq} \phi_\bq(\bG;\btau) e^{i(\bq+\bG)\cdot \br},
\end{equation}
where the $\bG$'s denote the reciprocal lattice vectors of the unit cell, and the $\bq$'s are the 
Bloch wavevectors commensurate with the BvK supercell.  The $\bq+\bG=0$ term is not included as it 
is canceled by the neutralizing background.  By combining Eqs.~\eqref{eq:fsv2} and 
\eqref{eq.varphi}-\eqref{eq.Vf} we find
\begin{equation}\label{eq.phi_with_K}
\phi_\bq(\bG;\btau) = \frac{e}{2\ve_0\epsilon_{\infty,2} N \Omega} \frac{e^{-i(\bqin+\bGin) \cdot 
\btau_\parallel} }{|\bqin+\bGin|} K(\bq+\bG,\tau_z),
\end{equation}
where $\Omega$ is the volume of the unit cell, and $N$ is the number of unit cells in the BvK 
supercell.

The next step in our procedure is to evaluate the potential of a point dipole. This is achieved 
simply by taking the linear variation of $\phi(\br;\btau)$ with respect to $\btau$: 
\begin{equation}
\frac{\D\phi(\br;\btau)}{\D \btau}\cdot \Delta \btau =\sum_{\bq,\bG\ne-\bq} e^{i(\bq+\bG)\cdot 
\br}\, \frac{\D \phi_\bq(\bG;\btau)}{\D \btau}\cdot \Delta \btau ,
\end{equation}
where $|\Delta \btau|$ is the dipole length. As in Ref.~\onlinecite{Verdi2015}, we now consider one 
such dipole potential for each atom in the BvK supercell: the charge will be given by the Born 
effective charge tensor, and the direction and length of the dipole will be given by the atomic 
displacement pattern $\Delta \btau^{(\bq\nu)}$ in a phonon mode with wavevector $\bq$:
\begin{equation}\label{eq.dVqnu}
\Delta V_{\bq\nu}(\bG) = -e\!\!\sum_{\k \bR, \a\b} \frac{\D \phi_\bq(\bG;\btau_{\k \bR})}{\D 
\tau_{\k \bR\a}} Z^*_{\k,\a\b} \Delta \tau_{\k\bR \b}^{(\bq\nu)}.
\end{equation}
The prefactor $-e$ has been added to obtain the potential energy experienced by an electron, and 
the atomic displacement in this expression is the same as in Eq.~\eqref{eq.disp}.

The electron-phonon matrix element $g_{mn\nu}(\bk,\bq) = \braket{\psi_{m\bk+\bq}| 
\Delta_{\bq\nu}V|\psi_{n\bk}}$ associated with the potential $\Delta V_{\bq\nu}$ is finally 
obtained by combining Eqs.~\eqref{eq.dVqnu}, \eqref{eq.phi_with_K}, \eqref{eq.disp}, and 
\eqref{eq.varphi_transl2}:
\begin{eqnarray} \label{eq:keyexpression}
&& g_{mn\nu}(\bk,\bq) = \frac{e^2}{2\ve_0\epsilon_{\infty,2}  \Omega} (\hbar/2 \w_{\bq\nu})^{1/2}\!\!  
\sum_{\bG\ne-\bq} \frac{\<u_{m\bk+\bq+\bG}| u_{n\bk}\>}{\left|\bqin\!+\!\bGin\right|} \nonumber \\ 
&& \,\,\times\sum_\k M_\k^{-1/2} e^{-i(\bqin+\bGin) \cdot \btau_{\k \parallel}} 
\sum_{\a\b}Z^*_{\k,\a\b} e_{\k\b,\nu}(\bq) \nonumber \\ && \,\,\times\left[\d_{\a,\parallel}\, i 
(\bq\!+\!\bG)_\a K(\bq\!+\!\bG,\tau_{\k z}) - \d_{\a,z}\, \frac{\D K(\bq\!+\!\bG,\tau_{\k z}) }{\D 
\tau_{\k z}} \right]. \nonumber \\\label{eq:frohlichquasi2d}
\end{eqnarray}
In this expression, we have taken into account the normalization $\psi_{n\bk} = 
N^{-1/2}e^{i\bk\cdot\br}u_{n\bk}$, where $u_{n\bk}$ is the Bloch-periodic part of the wavefunction, 
we used the perioidc gauge $\psi_{n\bk}=\psi_{n\bk+\bG}$, and the braket $\<\cdots\>$ integral is 
now performed over the unit cell of the stack.  Equation~\eqref{eq:frohlichquasi2d} is the central 
result of this work. It constitutes the generalization of the \textit{ab initio} \F\ 
electron-phonon matrix element, derived in  Ref.~\onlinecite{Verdi2015} for bulk 3D crystals, to 
periodic superlattices where two slabs alternate.  Equation~\eqref{eq:frohlichquasi2d} can be used 
(1) to improve the Wannier-Fourier interpolation of the electron-phonon matrix elements in the case 
of 2D materials, as shown in Refs.~\onlinecite{Verdi2015, Sjakste2015} for the 3D case and (2) to 
derive realistic analytical models of \F\ interactions in 2D and quasi-2D systems. In the remainder 
of this work we discuss both applications.

\section{The limits of 3D bulk crystal, 2D slab in vacuum, and atomically-thin monolayer in 
vacuum}\label{sec.limits}

In this section we show how Eq.~\eqref{eq:frohlichquasi2d} correctly reduces to the result of 
Ref.~\onlinecite{Verdi2015} in the limit of a single layer ($D=0$), and to the result of 
Ref.~\onlinecite{Sohier2016} in the limit of a slab in vacuum ($D=\infty$ and $\epsilon_{\infty,2}=1$). This 
latter situation would correspond, for example, to a suspended MoS$_2$ monolayer. To demonstrate 
the flexibility of our approach, we also consider a third option, namely an atomically thin 
monolayer in vacuum, where all the atoms have the same $z$ coordinate; this is the case, for 
example, of monolayer h-BN. 

\subsection{The limit of a 3D bulk crystal}\label{sec.bulklimit}

The \F\ matrix element for a 3D extended crystal is obtained by setting the thickness of the 
secondary layer to zero, $D=0$, in Eq.~\eqref{eq:frohlichquasi2d}. Since this parameter enters 
Eq.~\eqref{eq:frohlichquasi2d} only via the kernel $K(\bQ,\tau_{z})$, we start by considering this 
kernel.

In the limit of small $D$ ($D\ll c$), we have $d=c$, $\eta = Q_\parallel c$ from 
Eq.~\eqref{eq.eta}, and $\gamma^+ = (\a-1)/\a$, $\gamma^- = \a/(\a-1)$. Using these relations 
inside in Eq.~\eqref{eq:kernel}, after some algebraic manipulations we obtain
\begin{equation}
\lim_{D\rightarrow 0}K(\bQ,\tau_z) = 2\frac{\epsilon_{\infty,2}}{\epsilon_{\infty,1}} \frac{\Qin}{Q^2}e^{-iQ_z 
\tau_z}~.
\end{equation}
This expression can be replaced for the square brackets in Eq.~\eqref{eq:frohlichquasi2d}.  After 
this substitution, the matrix element for $D = 0$ reduces to
\begin{eqnarray}
 g_{mn\nu}(\bk,\bq) &=& i\frac{4\pi}{\Omega}\frac{e^2}{4\pi\ve_0} \sum_\k (\hbar/2 M_\k 
\w_{\bq\nu})^{1/2} \nonumber \\ &\times& \sum_{\bG\ne -\bq} e^{-i(\bq+\bG) \cdot \btau_{\k}} 
\<u_{m\bk+\bq+\bG}| u_{n\bk}\> \nonumber \\ & \times& \sum_{\a\b} \frac{ (\bq\!+\!\bG)_\a 
Z^*_{\k,\a\b} e_{\k\b,\nu}(\bq) }{\epsilon_{\infty,1}|\bq+\bG|^2}~. 
\end{eqnarray}
This result is essentially identical to the matrix element derived in Ref.~\onlinecite{Verdi2015} 
for bulk 3D crystals; see Eq.~(\ref{eq:verdi}). The only difference is that, to keep 
the derivation tractable, in the present study we have replaced the anisotropic dielectric 
permittivity tensor by a scalar isotropic permittivity. A similar choice was made in 
Ref.~\onlinecite{Sohier2016}.

In summary, the Fr\"ohlich matrix element given by Eq.~\eqref{eq:frohlichquasi2d} correctly reduces 
to the bulk limit in the case of a 3D homogeneous dielectric.

\subsection{The limit of an isolated 2D slab}\label{sec.sohier}

The other important limit to be investigated corresponds to the case of the \F\ interaction for an 
isolated 2D slab embedded in a dielectric continuum.  This limit is obtained by taking 
$D \gg d$ in Eq.~\eqref{eq:frohlichquasi2d}. The case of an isolated slab in vacuum is further 
obtained  by setting $\epsilon_{\infty,2} = 1$.

Taking the limit $D \gg d$ of the kernel function $K(\bQ,\tau_z)$ given in Eq.~\eqref{eq:kernel} 
requires a certain number of algebraic manipulations. Here we limit ourselves to remark that, in 
this limit,  $\gamma^+$ remains finite, $e^\eta$ scales as $e^{\Qin D}$, and $\gamma^-$ scales as 
$e^{2\Qin D}$. We find: 
\begin{eqnarray} 
&& \lim_{D\gg d} K(\bQ,\tau_z) = 2(\a-\b)\frac{\Qin}{\Qin^2+Q_z^2} \left\{  e^{-iQ_z\tau_z} + 
\right. \nonumber \\ &&\left. \hspace{0.4cm} + \frac{\b}{\a^2e^{2\Qin d}\!-\!\b^2} . \left[ e^{\Qin 
\tau_z} f_1(\bQ d) + e^{-\Qin\tau_z} f_2(\bQ d) \right] \vphantom{\frac{\b}{\b^2}}\!\!\right\}, 
\hspace{0.6cm} \label{eq.kern2d}
\end{eqnarray}
having defined:
\begin{eqnarray} 
f_1(\bQ d) &=& \a e^{2\Qin d}  + \b e^{(\Qin+iQ_z) d},\quad\\ f_2(\bQ d) &=& \b +\a e^{(\Qin+iQ_z) 
d}.  \label{eq.kern2df_2}
\end{eqnarray}
In this form, the effect of reduced dimensionality is not apparent, and the kernel is singular at 
long wavelength as for the bulk 3D case. In order to see the effect of 
dimensionality we need to carry out the summation over $G_z$ appearing in 
\eqref{eq:frohlichquasi2d}.  Since the overlap integral $\<u_{m\bk+\bq+\bG}| u_{n\bk}\>$ depends on 
$G_z$, we need to evaluate the sum
\begin{equation}\label{eq.sum}
\sum_{G_z} \<u_{m\bk+\bq+\bG}| u_{n\bk}\> K(\bQin,q_z+G_z,\tau_z).
\end{equation}
If the wave functions are localized within a characteristic length comparable to the thickness $d$ 
of the dielectric slab, and if we take the limit $D\gg d$, the overlap term $\<u_{m\bk+\bq+\bG}| 
u_{n\bk}\>$ becomes only weakly dependent on $G_z$, and the summation can be carried out explicitly. 
The specific details of the wave function localization around the slab are not critical to the final 
result, but in order to make contact with Ref.~\onlinecite{Sohier2016} we follow their choice and 
we set the Bloch-periodic components of the wave functions to be normalized rectangular functions in 
the direction perpendicular to the slab:
\begin{equation} \label{eq.rectbloch}
u_{n\bk}(\br) = \begin{cases} \sqrt{c/ \Omega d} & -d<z<0~,\\ 0 & 0<z<D~.  \end{cases}
\end{equation}
The corresponding wave functions are products of plane waves and this rectangular function.  This 
choice is legitimate since we are interested in the long-wavelength limit of the Fr\"ohlich matrix 
element, therefore the details of the wave function at the atomic scale do not matter. With the 
above choice the overlap integral becomes
\begin{equation}\label{eq.rect}
\<u_{m\bk+\bq+\bG}| u_{n\bk}\> = \delta_{mn}\d_{\bGin,0} \frac{1-e^{-iG_z d}}{i G_z d}.
\end{equation}
The summation in Eq.~\eqref{eq.sum} can now be carried out explicitly by combining 
Eqs.~\eqref{eq.kern2d} and \eqref{eq.rect}. Since we are interested in the limit $D\gg d$, we have 
that $q_z \rightarrow 0$, $G_z$ becomes a continuous variable, and the summation over $G_z$ can be 
replaced by an integral.  After some algebra we find
\begin{eqnarray}
&& \hspace{-1cm}\lim_{D\gg d}  \sum_{G_z} \<u_{m\bk+\bq+\bG}| u_{n\bk}\> K(\bQin,q_z+G_z,\tau_z) = 
\nonumber \\ &=&  \delta_{mn}\delta_{\bGin,0} \frac{c(\alpha -\beta)}{\Qin d} \bigg[  2 - e^{\Qin 
\tau_z} - e^{-\Qin \tau_z - \Qin d} \nonumber \\ &+&  \frac{\beta(1-e^{-\Qin d})}{ \alpha e^{\Qin 
d} -\beta } (e^{\Qin \tau_z}e^{\Qin d}+e^{-\Qin \tau_z})  \bigg]~.\label{eq.Gzkernel}
\end{eqnarray}
The matrix element given in Ref.~\onlinecite{Sohier2016} was obtained by considering that the ionic 
polarization is distributed uniformly across the slab. Their choice can be incorporated in the 
present formalism by averaging the atomic positions $\tau_z$ over the slab thickness.  To this aim, 
we define the kernel average as follows:
\begin{equation}
\< K(\bQin,q_z+G_z) \> = \frac{1}{d}\!\int_{-d}^0\!d\tau_z\, K(\bQin,q_z+G_z,\tau_z)
\end{equation}
By combining the last two equations and carrying out the integrals of the terms $e^{\pm\Qin 
\tau_z}$, we obtain
\begin{eqnarray}
&& \lim_{D\gg d}  \sum_{G_z} \<u_{m\bk+\bq+\bG}| u_{n\bk}\> \<K(\bQin,q_z+G_z)\> = \nonumber \\ 
&&\,\,\,=  \delta_{mn}\delta_{\bGin,0} c \frac{2 (2\alpha -1)}{\Qin d} \bigg[  1 + \frac{1}{\Qin d} 
\frac{ (2\alpha-1) (e^{\Qin d} -1) }{ 1-\alpha (1+e^{\Qin d}) } \bigg]. \nonumber \\
\end{eqnarray}
Now we can replace this expression inside Eq.~\eqref{eq:frohlichquasi2d}. We set 
$\epsilon_{\infty,1}=\epsilon_{\infty}$, $\epsilon_{\infty,2}=1$, to find an expression for 
the Fr\"ohlich 2D matrix element that is almost identical to the result of Ref.~\onlinecite{Sohier2016} 
as reported in Eq.~\eqref{eq:sohier} of the present work:
\begin{eqnarray}
&& g_{mn\nu}(\bqin) = i\d_{mn}\frac{2\pi}{A}\frac{e^2}{4\pi\ve_0} \nonumber \\ && \qquad 
\times\sum_{\k,\a=\parallel,\b} \frac{q_\a}{\qin} Z^*_{\k,\a\b} \sqrt{\frac{\hbar}{2 
M_k\w_{\bq\nu}}} e^{-i\bqin \cdot \btau_{\k \parallel}} e_{\k\b,\nu}(\bqin) \nonumber \\ && \qquad 
\times \frac{1}{\epsilon_{\infty}}\frac{2}{\qin d} \bigg[  1 + \frac{\epsilon_{\infty}^{-1}}
{\qin d} \frac{ e^{\qin d} -1 }{ 1-(1+\epsilon_{\infty}^{-1})(1+e^{\qin d})/2 } \bigg]~.  
\label{eq.almostsohier}
\end{eqnarray}
Note that this expression contains only $\bqin$ because we are in the limit $q_z \rightarrow 0$, 
and it no longer depends on the electron wave vector $\bk$.

The equivalence between Eq.~\eqref{eq.almostsohier} and Eq.~\eqref{eq:sohier} can be made 
more apparent by introducing the mass-weighted mode-effective charge $Z_{\bqin \nu}^*$ as follows:
\begin{equation}
Z_{\bqin \nu}^* =   \sum_\k \sqrt{\frac{M_0}{M_k}} \,\widehat{\bq}_\parallel \cdot {\bf 
Z}^*_{\k}\cdot {\bf e}_{\k,\nu}(\bqin) e^{-i\bqin \cdot \btau_{\k \parallel}}, 
\end{equation}
where $\widehat{\bq}_\parallel$ is the unit vector in the direction of $\bqin$.  Using this 
definition, Eq.~\eqref{eq.almostsohier} can be rewritten more compactly as:
\begin{eqnarray}
g_{mn\nu}(\bqin) &=& i\d_{mn}\frac{2\pi}{A}\frac{e^2}{4\pi\ve_0} \sqrt{\frac{\hbar}{2 
M_0\w_{\bq\nu}}}\,Z_{\bqin \nu}^* \frac{1}{\epsilon_{\infty}}\frac{2}{\qin d} \nonumber \\ 
&\times & \bigg[  1 + \frac{\epsilon_{\infty}^{-1}}{\qin d} \frac{ e^{\qin d} -1 }
{ 1-(1+\epsilon_{\infty}^{-1})(1+e^{\qin d})/2 } \bigg].\quad \label{eq.almostsohier2}
\end{eqnarray}
Equation~\eqref{eq:sohier} is recovered by taking the $\bqin \rightarrow 0$ limit of $Z_{\bqin 
\nu}^*/\w_{\bq\nu}^{1/2}$, and by adding the square moduli of the matrix elements for the two 
in-plane directions of the zone-center longitudinal optical modes. 

\subsection{The limit of an isolated 2D monolayer where all atoms have the same $z$-coordinate} 
\label{sec:model}

An alternative expression to the 2D Frh\"olich matrix element can be obtained in the limit of 
materials consisting of a single atomic layer where all atoms have the same $\tau_{z}$, such as for 
example monolayer boron nitride.

In this scenario we can go back to Eq.~\eqref{eq.Gzkernel}, and instead of averaging $\tau_z$ over 
the dielectric slab (which is equivalent to the approach of Ref.~\citenum{Sohier2016} as shown in 
Sec.~\ref{sec.sohier}), we can simply set $\tau_z = -d/2$ for all atoms, which corresponds to 
having the monolayer in the middle of the dielectric slab. We find:
\begin{eqnarray}
&&\lim_{D\gg d}  \sum_{G_z} \<u_{m\bk+\bq+\bG}| u_{n\bk}\> K(\bQin,q_z+G_z,\tau_z = -d/2) = 
\nonumber \\ &&\hspace{15pt}=  \delta_{mn}\delta_{\bGin,0} \frac{ 2(2\alpha -1)c}{\Qin d} \bigg[ 1+ 
\frac{ (2\alpha  - 1 ) e^{\Qin d/2} }{ 1-\alpha (e^{\Qin d}+1)  } \bigg]~.
\end{eqnarray}
By repeating the same steps that led to Eq.~\eqref{eq.almostsohier} in the slab case, we obtain the 
Fr\"ohlich matrix element for an isolated 2D monolayer: 
\begin{eqnarray}
g_{mn\nu}(\bqin) &=& i\d_{mn}\frac{2\pi}{A}\frac{e^2}{4\pi\ve_0} \sqrt{\frac{\hbar}{2 
M_0\w_{\bq\nu}}}\,Z_{\bqin \nu}^* \frac{1}{\epsilon_{\infty}}\frac{2}{\qin d} \nonumber \\ 
&\times & \bigg[ 1+ \frac{ \epsilon_{\infty}^{-1} e^{\qin d/2} }
{ 1-(1+\epsilon_{\infty}^{-1})(1+e^{\qin d})/2  } \bigg].\quad  
\label{eq.monolayer}
\end{eqnarray}
In the limit of long wavelengths, this expression reduces to the simplified form:
\begin{equation}\label{eq.2d-simpl}
\lim_{\qin \rightarrow 0}g_{mn\nu}(\bqin) = i\d_{mn}\frac{2\pi}{A}\frac{e^2}{4\pi\ve_0} 
\sqrt{\frac{\hbar}{2 M_0\w_{\bq\nu}}}\,Z_{\bqin \nu}^* \, \frac{1}{1+\qin/q_0}, 
\end{equation}
having defined:
\begin{equation}\label{eq.q0}
q_0 = \frac{4 \epsilon_{\infty}}{2\epsilon_{\infty}^2 -1}\frac{1}{d}.
\end{equation}
This approximation to Eq.~\eqref{eq.monolayer} remains very close to the the original equation 
through the entire range of wave vectors $\qin$, therefore this simplified matrix element is 
especially useful to derive analytic expressions for the Frh\"ohlich coupling in two dimensions.

\section{Model matrix elements without Born effective charges}\label{sec.simplemodel}

The 2D \F\ matrix element for isolated slabs and monolayers as derived in 
Eqs.~\eqref{eq.almostsohier2}, \eqref{eq.monolayer}, and \eqref{eq.2d-simpl}, can be simplified 
further by expressing the mode effective Born charges in terms of the dielectric constants of the slab. This 
step is useful to obtain the 2D analog of the \F\ matrix element used for bulk 2D solids, 
Eq.~\eqref{eq.fr.model5}.

In the long-wavelength limit, the relation between dielectric constants and the mode effective 
charges is provided by Eq.~(56) of Ref.~\citenum{Gonze1997}, here rewritten without assuming 
Hartree units (in the following we use $\epsilon_\infty,\epsilon_0$ or $\epsilon^\infty,\epsilon^0$ 
interchangeably to accommodate the other indices as needed):
\begin{equation}\label{eq.gonze1}
\epsilon_{\widehat{\bq}_\parallel}^0 - \epsilon_{\widehat{\bq}_\parallel}^\infty = \frac{e^2}{4\pi 
\varepsilon_0}\frac{4\pi}{\Omega}\sum_\nu \frac{(Z_{\widehat{\bq}_\parallel, 
\nu}^*)^2}{M_0\omega_{0,\nu}^2}.
\end{equation}
In this expression, $\epsilon_{\widehat{\bq}_\parallel}^0$ and 
$\epsilon_{\widehat{\bq}_\parallel}^\infty$ denote the relative static and high-frequency 
dielectric permittivities evaluated along the direction $\widehat{\bq}_\parallel$, and $\w_{0,\nu}$ 
indicates the frequency of the vibrational mode, $\nu$, at $\bq =0$. This frequency \textit{does 
not} include the nonanalytic part of the dynamical matrix, i.e., it is the TO frequency.

To make contact with the standard \F\ model for 3D bulk systems, we must convert the TO frequencies 
in Eq.~\eqref{eq.gonze1} into LO frequencies. This can be achieved via the generalized 
Lyddane-Sachs-Teller relations, namely, Eq.~(64) of Ref.~\citenum{Gonze1997}:
\begin{equation}\label{eq.gonze2}
\prod_\nu \frac{\w_{\bqin\rightarrow 0,\nu}^{2}}{\w_{\bqin=0,\nu}^{2}} = 
\frac{\epsilon_{\widehat{\bq}_\parallel}^0} {\epsilon_{\widehat{\bq}_\parallel}^\infty}.
\end{equation}
The frequencies in the denominator of this expression are the TO frequencies, and those in the 
numerator are the LO frequencies. By combining Eqs.~\eqref{eq.gonze1} and \eqref{eq.gonze2}, and 
considering a single infrared-active mode, we find
\begin{equation}\label{eq.born}
\frac{(Z_{\widehat{\bq}_\parallel, \nu}^*)^2}{\w_{\bqin\rightarrow 0,\nu}} = \frac{\varepsilon_0 
\Omega M_0}{e^2} \, \w_{\bqin\rightarrow 0,\nu} \,(\epsilon_{\widehat{\bq}_\parallel}^{\infty})^2 
\left( \frac{1}{\epsilon_{\widehat{\bq}_\parallel}^\infty}-\frac{1} 
{\epsilon_{\widehat{\bq}_\parallel}^0}\right).
\end{equation}
This relation should be replaced inside Eqs.~\eqref{eq.almostsohier2}, \eqref{eq.monolayer}, and 
\eqref{eq.2d-simpl}, after noting that the frequency $\w_{\bq\nu}$ appearing in those expressions 
corresponds to $\w_{\bqin\rightarrow 0,\nu}$, i.e., the LO frequency.

The dielectric constants in Eq.~\eqref{eq.born} correspond to the entire supercell. In order to 
disentangle the screening by the dielectric slab and by the environment, we use the standard rule 
for a stack of dielectrics:
\begin{equation}\label{eq.diel}
d \,\epsilon_{1} + D\, \epsilon_{2} = (d+D) \,\epsilon_{\widehat{\bq}_\parallel}.
\end{equation}
After replacing these expression in Eq.~\eqref{eq.born}, taking the limit $D\gg d$, and setting 
$\epsilon_1=\epsilon$ and $\epsilon_2=1$, we find
\begin{equation}\label{eq.gonze4}
\frac{(Z_{\widehat{\bq}_\parallel, \nu}^*)^2}{\w_{\bqin\rightarrow 0,\nu}} = \frac{\varepsilon_0 Ad 
M_0}{e^2} \, \w_{\rm LO} (\epsilon_0- \epsilon_\infty),
\end{equation}
where the LO frequency is given by $\w_{\bqin\rightarrow 0,\nu}=\w_{\rm LO}$.  The left-hand side 
is now expressed in terms of intrinsic properties of the slab, and does not depend on the size of 
the vacuum buffer. We also note that the Born charge evaluated along the direction parallel to the 
slab does not depend on the size $c$ of the supercell.\citep{Giustino2005prl}

Incidentally, we remark that in the limit of $D\gg d$ the LO and TO frequencies tend to the same 
value, because the dielectric constant of the vacuum buffer (or any uniform dielectric buffer) 
overwhelms the dielectric screening of the slab. This is easily proven by replacing 
Eq.~\eqref{eq.diel} inside Eq.~\eqref{eq.gonze2} and taking the limit $D\gg d$. This observation is 
in agreement with the absence of LO-TO splitting in 2D materials discussed in 
Ref.~\onlinecite{Sohier2017}.

Using Eq~\eqref{eq.gonze4}, we can now rewrite Eqs.~\eqref{eq.almostsohier2}, \eqref{eq.monolayer}, 
and \eqref{eq.2d-simpl} without resorting to the Born charges:
\begin{equation}
g_{mn\nu}(\bqin) = i\d_{mn}\left[\frac{\pi}{2}\frac{e^2}{4\pi\ve_0} \frac{d }{A} \, \hbar\w_{\rm LO} 
(\epsilon_0- \epsilon_\infty)\right]^{1/2} \!\!\!\!\!f(\qin d,\epsilon_\infty),
\end{equation}
where the dimensionless function $f(\qin d,\epsilon_\infty)$ depends on the chosen approximation 
for the slab. The function corresponding to the model of Ref.~\citenum{Sohier2016} that yields 
Eq.~\eqref{eq.almostsohier2} is
\begin{eqnarray}
f_1(\qin d,\epsilon^\infty) &=& \frac{1}{\epsilon_\infty}\frac{2}{\qin d} \bigg[  1 + 
\frac{\epsilon_\infty^{-1}}{\qin d} \times \nonumber\\ &\times&\frac{ e^{\qin d} -1 }{ 
1-(1+\epsilon_\infty^{-1})(1+e^{\qin d})/2 } \bigg].
\end{eqnarray}
The function corresponding to the present monolayer model that yields Eq.~\eqref{eq.monolayer} is
\begin{equation}
f_2(\qin d,\epsilon_\infty) = \frac{1}{\epsilon_\infty}\frac{2}{\qin d} \bigg[ 1+ \frac{ 
\epsilon_\infty^{-1} e^{\qin d/2} }{ 1-(1+\epsilon_\infty^{-1})(1+e^{\qin d})/2  } \bigg],
\end{equation}
and lastly the function corresponding to the long-wavelength limit of the monolayer model, yielding 
Eq.~\eqref{eq.2d-simpl}, is
\begin{equation}\label{eq.f3}
f_3(\qin d,\epsilon_\infty) = \frac{1}{1+\qin/q_0},
\end{equation}
with $q_0$ given by Eq.~\eqref{eq.q0}.

These expressions can be used to study \F\ interactions in two dimensions without performing 
explicit \textit{ab initio} calculations. For the reader's convenience, we quote in full the 
expression corresponding to the simplest approximation, Eq.~\eqref{eq.f3}:
\begin{equation}\label{eq.simplemodel}
g_{mn\nu}(\bqin) = i\d_{mn}\left[\frac{\pi}{2}\frac{e^2}{4\pi\ve_0} \frac{d }{A} \, \hbar\w_{\rm LO} 
(\epsilon_0- \epsilon_\infty)\right]^{1/2}\!\!  \frac{q_0}{q_0+\qin},
 \end{equation}
where $q_0$ is given by Eq.~\eqref{eq.q0}, which we reproduce here for convenience:
\begin{equation}\label{eq.q0b}
q_0 = \frac{4 \epsilon_{\infty}}{2\epsilon_{\infty}^2 -1}\frac{1}{d}.
\end{equation}
The key difference between this expression and the 3D \F\ matrix element in 
Eqs.~\eqref{eq.fr.model5} and \eqref{eq:couplingstrength3d} is that the limit $\bqin\rightarrow 0$ 
is finite, as expected. 

\section{Methods and Results} \label{sec:results}

\subsection{Computational details} \label{sec:computational_details}

In order to demonstrate the method outlined in 
Secs.~\ref{sec:theory}-\ref{sec.simplemodel}, we consider hexagonal boron nitride (h-BN) and 
molybdenum disulfide (MoS$_2$) as test systems.  Both compounds crystallize in a layered hexagonal 
structure with space group P6$_3$/mmc.  We performed calculations of the ground state electronic 
structure and phonon dispersions using density-functional theory and density-functional 
perturbation theory, using plane waves and pseudopotentials, as implemented in the \Call{Quantum 
ESPRESSO}{} materials simulation suite.\citep{Giannozzi2009,Giannozzi2017} We used ONCV 
pseudopotentials\citep{Hamann2013,Schlipf2015} (h-BN: PBE;\citep{Perdew1996} MoS$_2$: 
LDA\citep{Ceperley1980,Zunger1981}), with plane waves kinetic energy cutoffs of 125~Ry and 135~Ry, 
respectively.  In both cases, the Brillouin zone grid was sampled by using a $\Gamma$-centered 
uniform grid of 14$\times$14$\times$6 points.\citep{Lebedev2016} The lattice vectors and internal 
coordinates of the bulk crystals were optimized with this setup.  DFPT calculations of phonons and 
electron-phonon couping matrix elements were performed using both a periodic supercell geometry, 
and using 2D Coulomb truncation.\citep{Sohier2017prb} The interpolation of the matrix 
elements\citep{Giustino2007} was performed using the \Call{Wannier90}{} \citep{Mostofi2014} and 
\Call{EPW}{}\citep{Ponce2016} codes.

To build the monolayer models, we start from bulk crystals of h-BN and MoS$_2$, we remove one of 
the two layers in the crystalling unit cell, and we expand the vacuum gap in the $z$ direction to 
$c=20$~\AA. With this choice, the direct gap nature of the monolayers is correctly captured. Soft
phonons corresponding to the interlayer breathing mode are found for $8\times8\times1$ and $12\times12\times1$ 
Brillouin zone grids. To avoid these soft modes, we start from a coarser $4\times4\times1$ grid.
This choice, albeit approximate, does not affect the long-range \F\ component of the electron-phonon matrix element.
The in-plane lattice parameters were set to the values optimized the corresponding bulk crystals, 
namely, $a=2.51$~\AA\ for h-BN and $a=3.12$~\AA\ for MoS$_2$. In Table~\ref{tab:table1} we report 
the calculated structural parameters, band effective masses, Born charges, and phonon energies.  
These values agree with previous literature.\citep{Kormanyos2015,Ferreira2019}

\begin{table}
\center 
 \begin{tabular}{ llrrl}
 \hline 
 \hline\\[-8pt]
& \textrm{Property} & 
\textrm{Symbol} &  
\textrm{Value} & Unit \\[2pt]
 \hline \\[-8pt]
h-BN &Lattice constant&$a$ &  2.511 & \AA\ \\
&Aspect ratio&$c/a$ &  7.965 & \\
&Effective mass&$m^*_{h,\parallel}$ & 0.650 & $m_e$ \\
&Base area of unit cell&$A$ & 5.460 & \AA$^2$  \\
&Born charge &$Z^*_\parallel$(B) &    2.702 & $e$  \\
&&$Z^*_\parallel$(N) &  $-$2.702 & $e$  \\
&LO phonon energy&$\hbar \omega_{\text{LO}}$ &  181.560 & meV  \\
&&$\hbar \omega^\text{2D}_{\text{LO}}$ & 166.992 & meV \\
& Dielectric thickness & $d$ & 2.648 & \AA\ \\
& High-frequency permittivity & $\epsilon_\infty$ & 5.695\\
& Low-frequency permittivity & $\epsilon_0$ & 7.921 \\[2pt] 
\hline\\[-8pt]
MoS$_2$ & Lattice constant&$a$ & 3.123 & \AA\  \\
& Aspect ratio&$c/a$ &  6.404 & \\
& Effective mass&$m^*_{h,\parallel}$ & 0.570 &$m_e$  \\
& Base area of unit cell&$A$ & 8.446  & \AA$^2$ \\
& Born charge &$Z^*_\parallel$(Mo) &    $-$1.170 & $e$  \\
& &$Z^*_\parallel$(S) &  0.585 &$e$  \\
& LO phonon energy&$\hbar \omega_{\text{LO}}$ &  48.734 &meV  \\
& &$\hbar \omega^\text{2D}_{\text{LO}}$ & 48.419 &meV \\
& Dielectric thickness & $d$ & 5.468 & \AA\ \\
& High-frequency permittivity & $\epsilon_\infty$ & 16.424\\
& Low-frequency permittivity & $\epsilon_0$ & 16.672 \\[2pt] 
 \hline\hline
\end{tabular}
\caption{\label{tab:table1}  Calculated material parameters of monolayer h-BN and monolayer 
MoS$_2$. $m_e$ and $e$ are the electron mass and charge, respectively.  
$\omega^\text{2D}_{\text{LO}}$ is the phonon energy calculated using 2D Coulomb truncation. The 
calculation of the dielectric thickness $d$ and the effective dielectric constants 
$\epsilon^\infty$ and $\epsilon^0$ is discussed in Sec.~\ref{sec:validation}.  
 }
\end{table}

\subsection{Validation of the formalism against explicit DFPT calculations} \label{sec:validation}

In this section we validate our formulation of the Fr\"ohlich matrix element in a slab geometry, by 
comparing our expression~(\ref{eq:keyexpression}) to explicit DFPT calculations.  We consider 
two types of calculations: (1) DFPT calculations in a periodic supercell configuration, and (2) 
DFPT calculations using Coulomb truncation, which are meant to describe an isolated monolayer 
without periodic images. We recall that Eq.~(\ref{eq:keyexpression}) is a completely general 
expression that should be able to reproduce both of these scenarios.

The kernel appearing in Eq.~(\ref{eq:keyexpression}) and reported in Eq.~\eqref{eq.fullkernel} 
depends on the values $\epsilon_{\infty,1}$, $\epsilon_{\infty,2}$, and $d$ that define 
the dielectric profile; see Eq.~\eqref{eq:profile}.  To extract these quantities from DFPT calculations, 
we use a simple capacitor stack model following Refs.~\onlinecite{Freysoldt2008} and \onlinecite{Laturia2018}. 
The dielectric constant of the supercell can be written in terms of the dielectric constants of the 
slab and the dielectric environment as
\begin{equation} \label{eq:capacitor_parallel}
c \,\epsilon_\parallel^{\rm sc} = D\epsilon_\parallel^{\rm env} + d\,\epsilon_\parallel^{\rm slab},
\end{equation}
\begin{equation} \label{eq:capacitor_parallel2}
c \,\epsilon_{\perp,\rm sc}^{-1} = D\epsilon_{\perp,\rm env}^{-1} + d\,\epsilon_{\perp,\rm 
slab}^{-1}.
\end{equation}
In these expressions, ``sc'' refers to the supercell, ``env'' stands for the environment (e.g.,
vacuum), and ``slab'' stands for the 2D layer; $\parallel$ and $\perp$ refer to the dielectric 
constants in the direction parallel and perpendicular to the layer, respectively. The values 
$\epsilon_\parallel^{\rm sc}$ and $\epsilon_\perp^{\rm sc}$ are obtained from DFPT calculations on 
the supercell, while the values $\epsilon_{\parallel,\rm slab}^{-1}$, $\epsilon_{\perp,\rm 
slab}^{-1}$, and $d$ need to be extracted from 
Eqs.~\eqref{eq:capacitor_parallel} and \eqref{eq:capacitor_parallel2}. In the following  we consider a 
vacuum buffer, so that $\epsilon_\parallel^{\rm env}=\epsilon_\perp^{\rm env}=1$. 

\begin{figure}
\centering
\includegraphics[width=1.02\columnwidth]{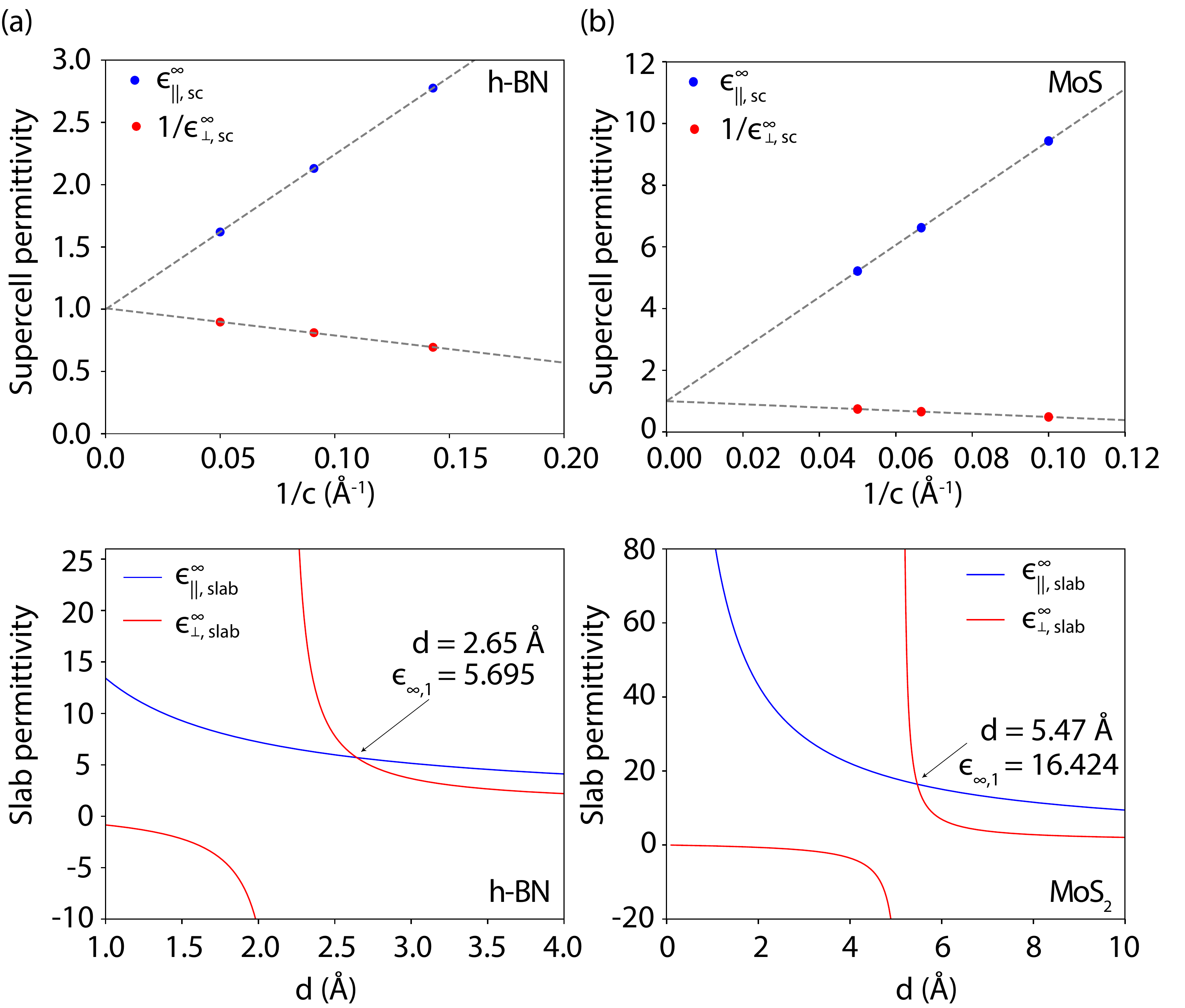}
\caption{ 
(a, b) Calculated high-frequency dielectric constants of h-BN (a) and 
MoS$_2$ (b) monolayers in a supercell geometry, as a function of cell length $c=d+D$ along the $z$ 
direction. See Fig.~\ref{fig:fig1} for details of the supercell construction. The dashed straight 
lines are guides to the eye and show that the supercell permittivity scale linearly with $1/c$. 
(c, d) Dielectric constants of monolayer h-BN (c) and MoS$_2$ (d), as extracted from 
Eqs.~\eqref{eq:capacitor_parallel} and \eqref{eq:capacitor_parallel2}, as a function of the dielectric 
thickness $d$. The crossing of the curves for the parallel and the perpendicular dielectric 
constants identify the effective dielectric thickness and permittivity of each slab. The symbols 
$\parallel$ and $\perp$ indicate dielectric constant parallel and perpendicular to the slab 
surface, respectively.}
\label{fig:fig2}
\end{figure}

To verify Eqs.~\eqref{eq:capacitor_parallel} and \eqref{eq:capacitor_parallel2}, in 
Figs.~\ref{fig:fig2}(a) and \ref{fig:fig2}(b) we plot the dielectric constants of supercells containing an h-BN 
(a) and a MoS$_2$ (b) monolayer as a function of the $c$ parameter. In agreement with the above 
equations, the dielectric constants of the supercell vary linearly with $c$. While 
Eqs.~\eqref{eq:capacitor_parallel} and \eqref{eq:capacitor_parallel2} do not uniquely define 
$\epsilon_{\parallel,\rm slab}^{-1}$ and $\epsilon_{\perp,\rm slab}^{-1}$, we can introduce one 
additional relation to be consistent with the assumption of isotropic permittivity used in 
Eq.~\eqref{eq:profile}:
\begin{equation}\label{eq.iso}
\epsilon_{\perp,\rm slab} = \epsilon_{\parallel,\rm slab}.
\end{equation}
This relation is justified on the grounds that our matrix elements is designed to capture the 
long-range behavior of the Fr\"ohlich interaction, as already discussed in 
Ref.~\onlinecite{Sohier2016}.  Taken together, Eqs.~\eqref{eq:capacitor_parallel}-\eqref{eq.iso} 
uniquely define the dielectric constants of the slab. A graphical solution of these equations is 
shown in Figs.~\ref{fig:fig2}(c) and \ref{fig:fig2}(d) for h-BN and MoS$_2$, respectively. The effective 
dielectric thickness of the slabs and the associated dielectric constants are reported in 
Table~\ref{tab:table1}.

\begin{figure}
\centering
\includegraphics[width=1.06\columnwidth]{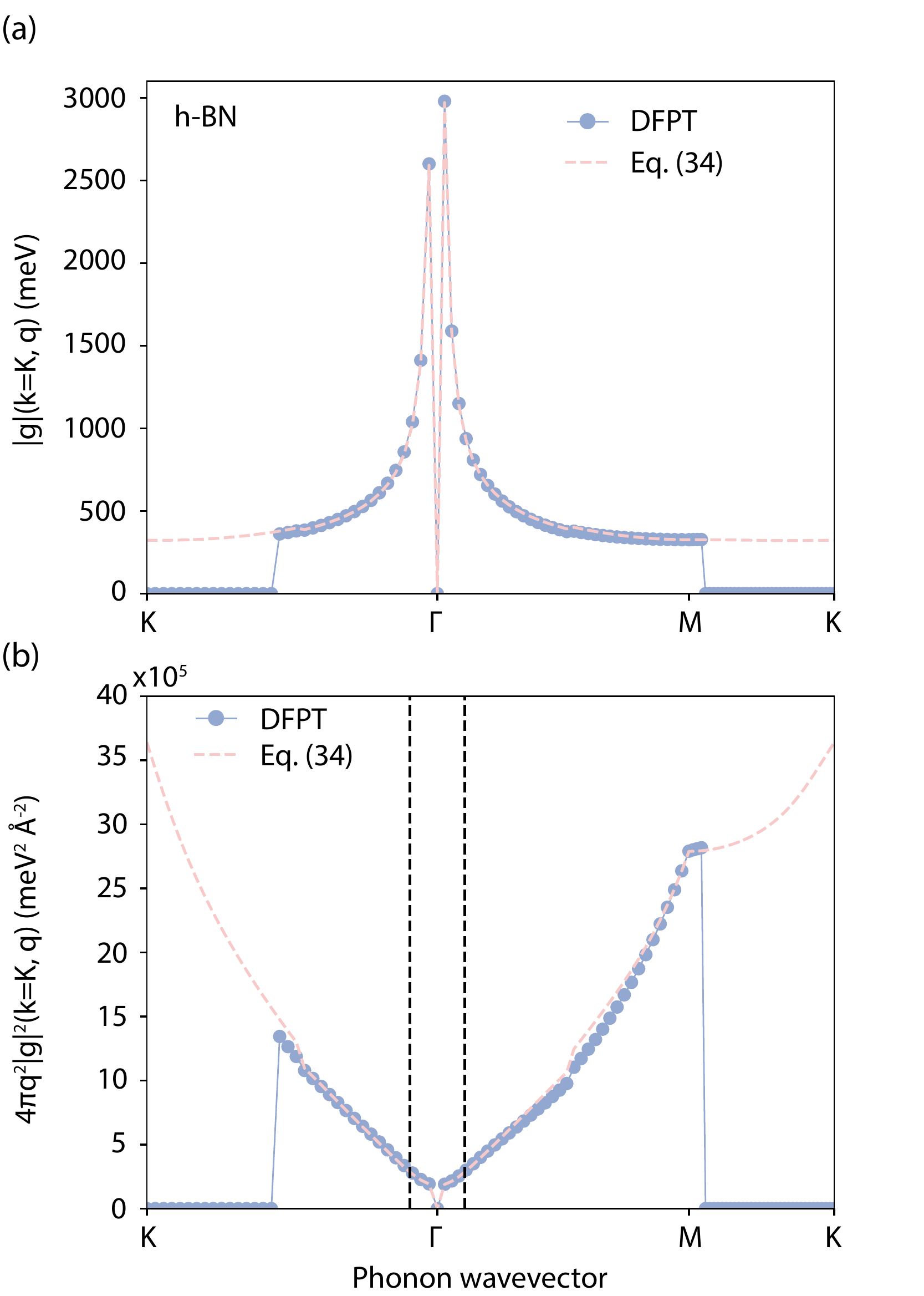}
\caption{
(a) Electron-phonon matrix elements of monolayer h-BN calculated along a high-symmetry 
path in the 2D Brillouin zone. We report the results of explicit DPFT calculations
(blue disks) and calculations using the matrix element in Eq.~(\ref{eq:keyexpression}) (pink lines).
In both cases we show the average of $|g_{mn\nu}(\bk,\bq_\parallel)|^2$ over the TO and LO modes to
avoid the discontinuity resulting from crossing phonon bands, $|g| = \left[{\sum}_{\nu}|g_{mn\nu}(\bk,\bq)|^2 
\right]^{1/2}$, for 
$m,n,\bk$ corresponding to the valence band maximum at the $K$ point. The calculations were performed by using a supercell with a monolayer of h-BN and
a vacuum buffer, with a total size $c=20$~\AA\ along the $z$ direction. (b) Same raw data as in (a),
but this time the matrix element is scaled by the phase-space volume element in three dimensions, 
$4\pi |\bq_\parallel|^2|g_{mn\nu}(\bk,\bq_\parallel)|^2$.
The vertical dashed lines indicate wavevectors such that $|\bq_\parallel|=\pi/c$.}
\label{fig:fig3}
\end{figure}

In Fig.~\ref{fig:fig3} we compare the \F\ matrix element obtained for the h-BN monolayer in two 
ways: (1) Using the present formalism, as expressed by Eq.~(\ref{eq:keyexpression}) (pink dash 
line) and (2) Using explicitly DFPT calculations in a periodic supercell \textit{without} Coulomb 
truncation (light purple discs). The unit cell size along the $z$ direction in these calculations 
is $d=20$~\AA. In Fig.~\ref{fig:fig3}(a) we show the modulus of the matrix element, 
$|g_{mn\nu}(\bk,\bq_\parallel)|$, for $m,n,\bk$ corresponding to the valence band maximum at the 
$K$ point, $\nu$ corresponding to the LO mode, and $\bq_\parallel$ along a high-symmetry path.  
We see that our formalism matches the explicit DFPT calculations near $\bq=0$, thereby confirming 
the validity of our approach. Our \F\ matrix element also matches DPFT calculations away from the 
zone center, which indicates that the interaction of the highest optical mode in monolayer h-BN 
system is \F-like in a large portion of the Brillouin zone. 

Figure~\ref{fig:fig3}(a) also shows that, although the supercell size in the $z$ direction is as 
large as $c=20$~\AA, the matrix element preserves the signature of 3D \F\ coupling, 
as it can be seen from the near-singular behavior for $\bq$ approaching the zone center. 
This effect is best visualized by considering the square modulus of the matrix element scaled by the 
phase-space volume element in three-dimensions, $|g_{mn\nu}(\bk,\bq)|^2 d\bq$.
This is the relevant quantity in applications, because typical expressions
for electron self-energies, carrier mobilities, and superconducting gap function, as reported,
e.g.,  in Ref.~\onlinecite{Giustino2017}, all contain an integration of the type 
  \begin{equation}\label{eq.exampleint}
    {\sum}_\nu\int \!\!\frac{d\bq}{\Omega_{\rm BZ}} \,\,|g_{mn\nu}(\bk,\bq)|^2 f_{mn\nu}(\bk,\bq),
  \end{equation}
where $\Omega_{\rm BZ}$ is the volume of the Brillouin zone, and the function $f_{mn\nu}(\bk,\bq)$ depends 
on the specific application. Figure~\ref{fig:fig3}(b) shows that $|g_{mn\nu}(\bk,\bq)|^2 d\bq$ scales
as $|\bq_\parallel|^{1}$ for $|\bq_\parallel|>\pi/c$, and tends to a constant value for $|\bq_\parallel|< \pi/c$. 
The implication is that, for $|\bq_\parallel|>\pi/c$ the coupling is markedly different from the standard 3D
\F\ interaction, while for $|\bq_\parallel|<\pi/c$ the coupling is of \F\ type and diverges as 
$|\bq_\parallel|^{-1}$. This singularity is a remnant of the \F\ interaction in three dimensions, and originates 
from the periodic images of the atomic dipoles along the $z$ direction. 
Indeed, to phonons with wavelengths longer than $c$, the supercell 
appears as a uniform material, for which the standard 3D \F\ interaction applies. In line with
this residual 3D-type interaction, we find a small but nonvanishing LO-TO splitting in the phonon
dispersion relations [$\hbar(\w_{\rm LO}-\w_{\rm TO})=15$~meV], whereas it is known that for 
a 2D system in isolation such a splitting is forbidden at the zone center.\citep{Sohier2017} 

The take-home message from Fig.~\ref{fig:fig3} is that Eq.~(\ref{eq:keyexpression}) correctly 
reproduces the \F\ matrix element in quasi-2D systems consisting of slab/vacuum stacks within
periodic BvK boundary conditions. Therefore our expression makes it possible to perform calculations
of electron-phonon interactions using Wannier interpolation~\citep{Ponce2016} as for 3D materials,
without requiring Coulomb truncation. 

\begin{figure}
\centering
\includegraphics[width=1.07\columnwidth]{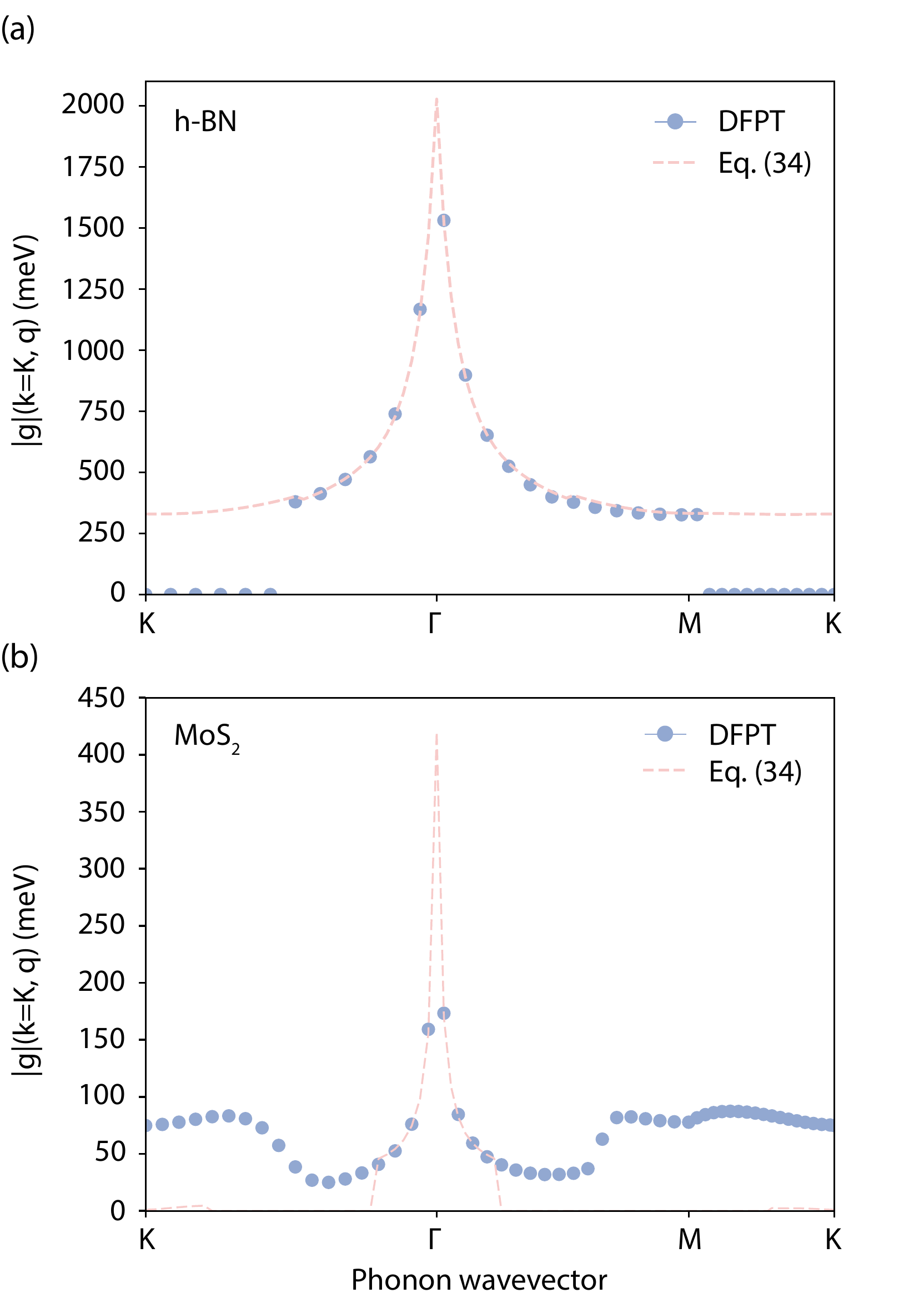}
\caption{ (a) Electron-phonon matrix elements of monolayer h-BN calculated along a high-symmetry 
path in the 2D Brillouin zone, for infinite interlayer separation. 
We report the results of explicit DPFT calculations employing 2D Coulomb truncation
(blue disks) and calculations using the matrix element in Eq.~(\ref{eq:keyexpression}) with 2D kernel [Eq.~(\ref{eq.kern2d})] 
(pink lines). In both cases we show the average of $|g_{mn\nu}(\bk,\bq_\parallel)|^2$ over the TO and LO modes to
avoid the discontinuity resulting from crossing phonon bands, $|g| = \left[{\sum}_{\nu}|g_{mn\nu}(\bk,\bq)|^2 
\right]^{1/2}$. In this expression, $m,n,\bk$ correspond to the valence band maximum at the $K$ point.
(b) Same as in (a), but for monolayer MoS$_2$ and infinite interlayer separation.
}
\label{fig:fig4}
\end{figure}

Now we move to the comparison between our formalism and DPFT calculations employing 2D
Coulomb truncation. In Fig.~\ref{fig:fig4} we compare the \F\ matrix element calculated for monolayer 
h-BN and MoS$_2$ via DFPT and Coulomb truncation\citep{Sohier2016} (blue disks) with our formalism 
(pink lines). In particular, we use the 2D kernel function Eq.~(\ref{eq.kern2d}), which corresponds to the $D\gg d$ 
limit of the exact matrix element in Eq.~(\ref{eq:keyexpression}), as discussed in Sec.~\ref{sec.sohier}. 
Figure.~\ref{fig:fig4}(a) shows the modulus of the matrix element $|g_{mn\nu}(\bk,\bq_\parallel)|$ for 
monolayer h-BN, with $m,n,\bk$ set to the valence band maximum at the $K$ point, $\nu$ corresponding 
to the LO mode, and $\bq_\parallel$ along a high-symmetry path. Figure~\ref{fig:fig4}(b) shows the 
corresponding quantity for monolayer MoS$_2$, also for the top of the valence band at $K$.

In both cases, we see that our formalism in the $D\gg d$ limit correctly reproduces the results
of explicit 2D DFPT calculations. In particular, now that the size of the vacuum buffer tends to
infinity, our formalism yields a finite, nonsingular \F\ matrix element at the zone center,
in complete agreement with truncated DFPT calculations. The level of agreement that can be 
seen in Figs.~\ref{fig:fig3} and \ref{fig:fig4} demonstrates the accuracy of our approach, and
shows that our method works seamlessly for periodic supercell calculations with finite vacuum 
buffer and for truncated 2D calculations with infinite vacuum.

It might be worth to point out that our matrix element is designed to describe the long-wavelength
limit of $g_{mn\nu}(\bk,\bq_\parallel)$, therefore a deviation between our results and DFPT
calculations at large $\bq_\parallel$ in Figs.~\ref{fig:fig3} and \ref{fig:fig4} is expected.
This deviation merely indicates that, at large $\bq_\parallel$, the coupling mechanism is no longer
of \F\ type. To describe the matrix element accurately throughout the Brillouin zone, it is
sufficient to combine the present formalism with Wannier-Fourier interpolation, as already
demonstrated in Ref.~\onlinecite{Verdi2015}.

\begin{figure}
\centering
\includegraphics[width=1.08\columnwidth]{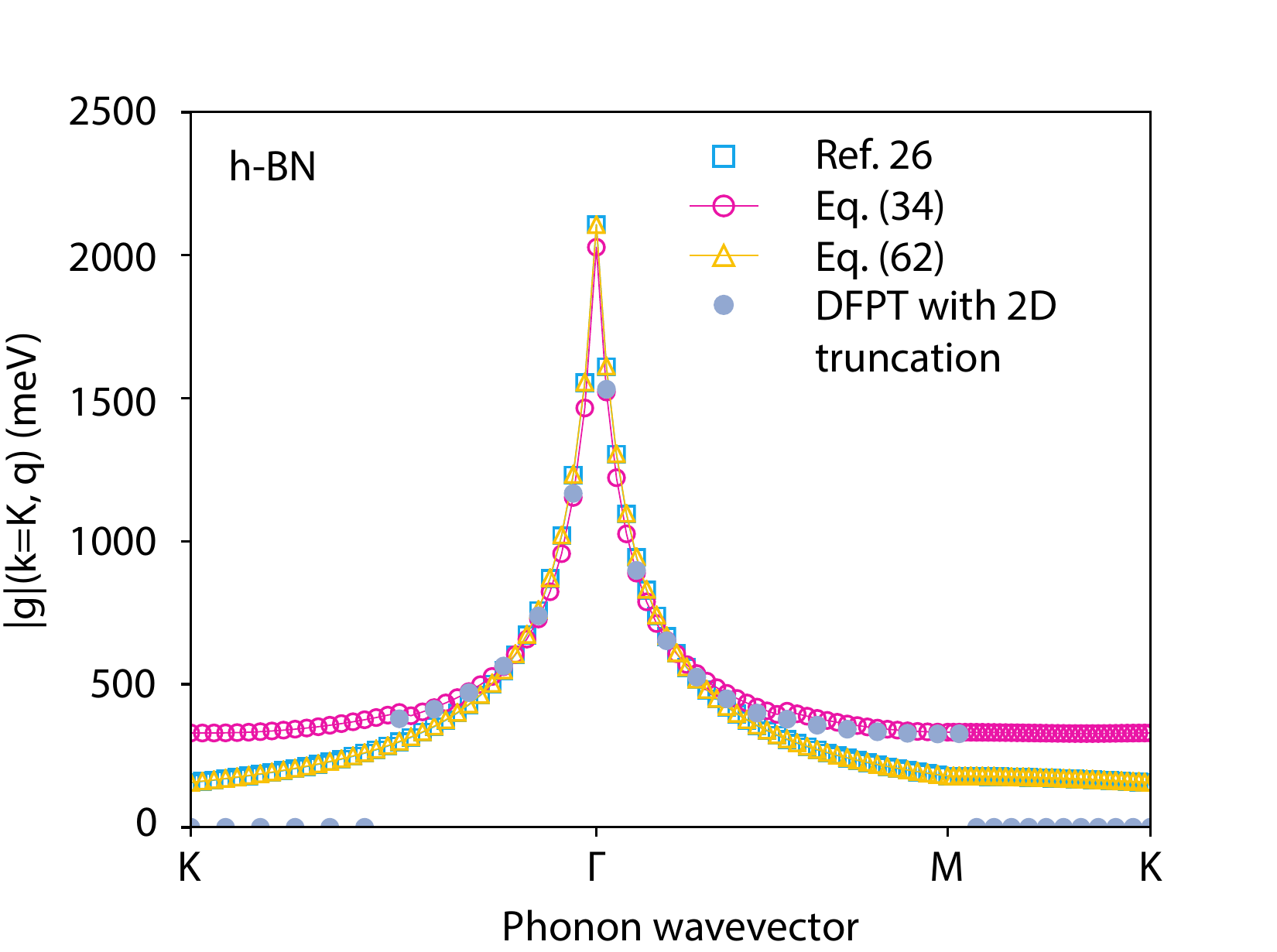}
\caption{Comparison between various models of the \F\ matrix elements for monolayer h-BN,
in the limit of infinite vacuum size. We show the modulus of the matrix element, $|g_{mn\nu}(\bk,\bq_\parallel)|$, 
for $m,n,\bk$ corresponding to the valence band maximum at the $K$ point, and $\nu$ corresponding to 
the LO mode. The gray disks are the reference DFPT calculations using 2D Coulomb truncation.
The data calculated using the method of Ref.~\onlinecite{Sohier2016} are shown as blue squares.
The data obtained with our exact matrix element Eq.~\eqref{eq:keyexpression} are in magenta.
The simplified model of Eq.~\eqref{eq.simplemodel}, using the parameters in Table~\ref{tab:table1},
is shown as yellow triangles.}
\label{fig:fig5}
\end{figure}

Next, we validate the simplified analytical model for 2D \F\ interactions given by Eq.~\eqref{eq.simplemodel}.
We recall that this model is useful to replace explicit DFPT calculations by a model matrix element
that depends only on macroscopic properties such as dielectric constants, dielectric thickness, and
phonon energy. Using the parameters in Table~\ref{tab:table1} for h-BN, we obtain the yellow triangles
in Fig.~\ref{fig:fig5}. These values are compared to the corresponding matrix elements according to
the method of Ref.~\onlinecite{Sohier2016} (blue squares), to our exact matrix element Eq.~\eqref{eq:keyexpression}
(magenta disks), and to DFPT calculations using 2D Coulomb truncation (gray disks).
It is apparent that, in the long-wavelength region, all these approaches are in very close agreement 
to each other. This successful comparison further demonstrates the validity of our approach, and
provides additional cross-validation of previously proposed approaches.\citep{Sohier2016} 

\subsection{Evolution of the \F\ coupling from 3D to 2D} \label{sec:model2dfrohlich}

In this section, we discuss the transition of the polar \F\ coupling from 3D to 2D using the
matrix element in Eq.~(\ref{eq:keyexpression}). To keep the focus on the essential physics,
we use the rectangular profile for the electron wave functions, as given by Eq.~(\ref{eq.rectbloch}),
and we employ materials parameter for h-BN, as reported in Table~\ref{tab:table1}.

\begin{figure}
\centering
\includegraphics[width=1.06\columnwidth]{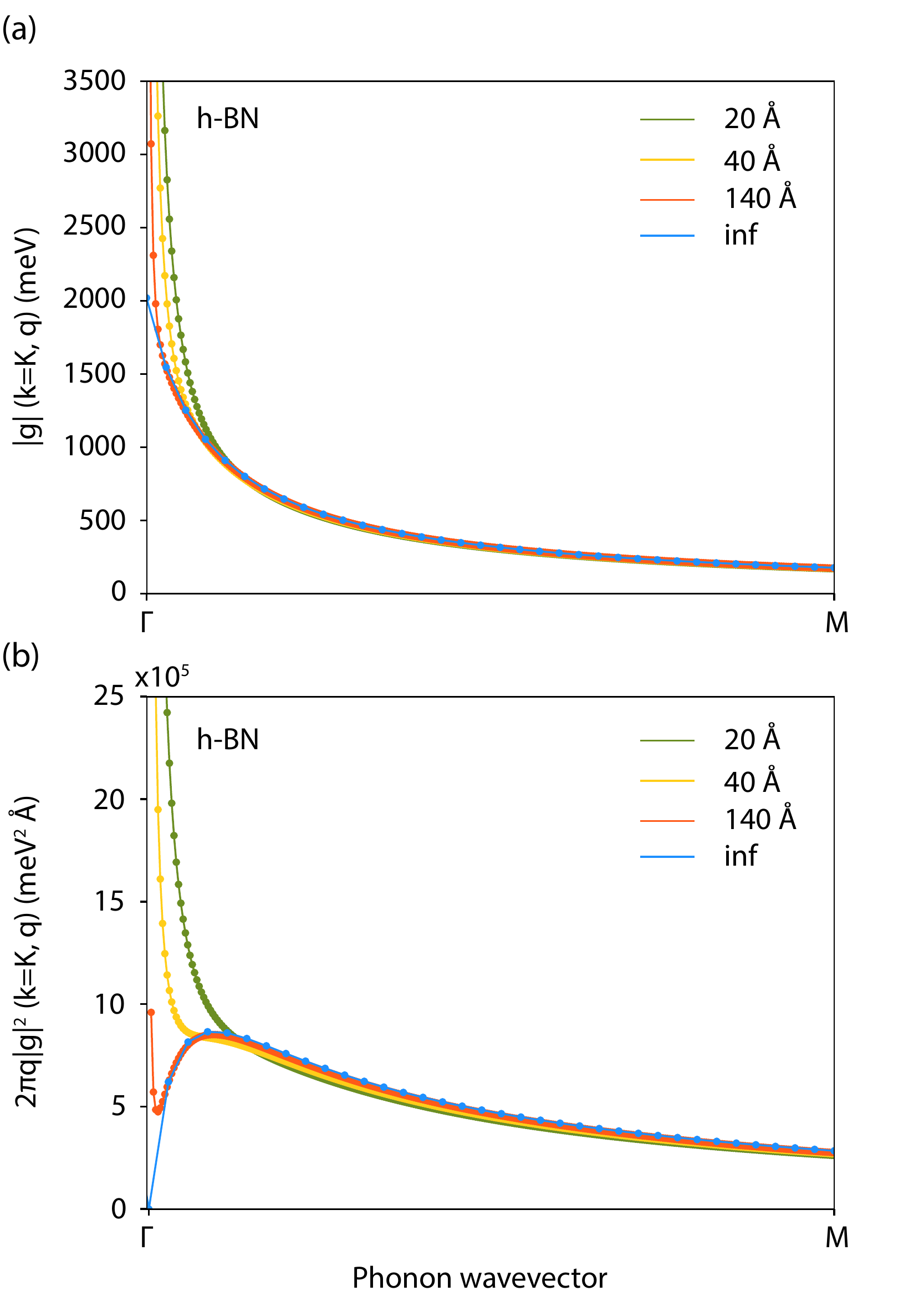}
\caption{
(a) Modulus of the long-range part of the electron-phonon matrix element, $|g_{mn\nu}(\bk,\bq_\parallel)|$, 
for $m,n,\nu$ corresponding to the valence band top of h-BN at the $K$ point and the LO mode. We consider phonon 
wave vectors along the $\Gamma M$ path. These data were calculated using Eq.~\eqref{eq:keyexpression},
for various supercell sizes $c$ along the $z$ direction: $c=20$~\AA~(green), 
40~\AA~(yellow), 140~\AA~(orange), and $c\rightarrow \infty$ (blue). 
(b) Same raw data as in (a), but this time plotted as $2\pi |\bq_\parallel||g_{mn\nu}(\bk,\bq_\parallel)|^2$.
}
\label{fig:fig6}
\end{figure}

Figure~\ref{fig:fig6}(a) shows the modulus of the electron-phonon matrix element, 
$|g_{mn\nu}(\bk,\bq_\parallel)|$, for $m,n,\nu$ corresponding to the valence band top of
h-BN at the $K$ point and the LO mode. We consider phonon wave vectors along the $\Gamma M$
path (the curves along the $\Gamma K$ path look very similar as already seen in 
Figs.~\ref{fig:fig3}-\ref{fig:fig5}). In Fig.~\ref{fig:fig6} we compare the matrix elements 
obtained for various supercell sizes $c$ along the $z$ direction, including $c=20$~\AA~(green), 
40~\AA~(yellow), 140~\AA~(orange), and $c\rightarrow \infty$ (blue). 

We can see that, for all finite values of $c$, the matrix element exhibits a singularity
at $\bq_\parallel = 0$, as in the case of the 3D \F\ interaction. See, for example, the curve
for $c=140$~\AA\ in Fig.~\ref{fig:fig6}(a). Although a nonsingular matrix element is only
obtained for $c\rightarrow \infty$, calculations using finite supercell sizes are still
meaningful, because what matters in actual calculations is the \textit{integral}
of the square modulus of the matrix element over the Brillouin zone, as already discussed
in relation to Eq.~\eqref{eq.exampleint}.
Figure~\ref{fig:fig6}(b) shows that this quantity converges
to the infinite-vacuum case with increasing $c$. Correspondingly, the singular region of the Brillouin
zone shrinks as $c$ increases, so that the contribution of the singularity to Eq.~\eqref{eq.exampleint} 
tends to become negligible at large $c$. Therefore, supercell calculations
\textit{without} Coulomb truncation constitute a viable strategy for studying \F\ interactions in 2D and quasi-2D
systems, with the proviso that the Wannier-Fourier interpolation strategy of Ref.~\onlinecite{Verdi2015}
be \textit{replaced} by the generalized interpolation procedure given by Eq.~\eqref{eq:keyexpression},
and that the convergence of the target physical property with respected to supercell size $c$ be achieved.

One further option that could be explored to accelerate the convergence of the calculations,
is to combine our matrix elements at finite $c$ with our expressions for $c\rightarrow\infty$.
For example, if we call $g^{\rm DFPT}$ the matrix element obtained by direct DFPT calculations
on a coarse Brillouin-zone grid, $g^{\rm 2D}(c)$ the matrix element obtained from 
Eq.~\eqref{eq:keyexpression} with a finite supercell size $c$, an $g^{\rm 2D}(c=\infty)$
the matrix element obtained from the limit form in Eq.~\eqref{eq.2d-simpl}, we could envision
a Wannier-Fourier interpolation strategy as follows: (1) Perform supercell calculations
without Coulomb truncation, yielding $g^{\rm DFPT}$. (2) Remove the long-range component
by subtracting $g^{\rm 2D}(c)$. This defines the short-range component $g^{\rm sr} = g^{\rm DFPT}- 
g^{\rm 2D}(c)$. (3) Interpolate the short-range component as usual.\citep{Giustino2007}
(4) Add to the interpolated short-range matrix elements the long-range component corresponding 
to the infinite-supercell limit,
$g = g^{\rm sr} +g^{\rm 2D}(c=\infty)$. This approach could also serve to test the convergence
of the calculations on the coarse grid vs. supercell size $c$.

\section{Conclusion} \label{sec:conclusion}

In this work we developed a unified description of the \textit{ab initio} Fr\"ohlich matrix element 
that enables calculations of long-range polar electron-phonon couplings in 3D and 
2D materials within a single formalism. We showed that the present approach recovers
the limits of bulk 3D materials and isolated 2D materials obtained in previous literature. In
particular, our generalized matrix element reduces to the 3D \F\ matrix element of Ref.~\onlinecite{Verdi2015}
when the interlayer separation $D$ between periodic images of the slab vanishes, and it reduces
to the 2D \F\ matrix element of Ref.~\onlinecite{Sohier2016} when the interlayer separation $D$
becomes infinite. 

We validated our methodology by performing DFPT calculations for two systems, monolayer h-BN and
monolayer MoS$_2$. In each case, we performed DFPT calculations using finite-size supercells
without Coulomb truncation, as well as DFPT calculations employing 2D Couloumb truncation.
In both cases our \F\ matrix element successfully matches explicit DFPT calculations in the
long-wavelength region. These results indicate that the present technique is ready to be employed in
conjunction with Wannier-Fourier interpolation of the electron-phonon matrix element.\citep{Giustino2007}
In particular, the present approach can be implemented as a straightforward extension
of the method of Ref.~\onlinecite{Verdi2015} in existing software packages like \Call{EPW}{}.\citep{Ponce2016}

In this work we also developed a minimal model of polar electron-phonon interactions in 2D.
In fact, Eq.~\eqref{eq.simplemodel} provides a simple yet very accurate expression for the
\F\ matrix element that depends only on the lattice parameters, the characteristic phonon
energy, and the static and high-frequency dielectric constants of the 2D material. Although
similar expressions were reported in previous literature,\citep{Kaasbjerg2012,Sohier2016}
this work establishes a transparent and direct link with macroscopic materials properties
that are readily available. We expect that this minimal model will facilitate the
investigation of \F\ couplings in 2D using model Hamiltonian approaches, and will help extracting
the essential physics from advanced \textit{ab initio} calculations.

We hope that the present cross-dimensional generalization of the \textit{ab initio} \F\ matrix element 
will enable further work in the physics of electron-phonon interactions in semiconductor/insulator interfaces, 
surfaces, 2D materials and their heterostructures.

\appendix

\section{Kernel function}\label{app.kernel}

In this Appendix we provide the complete expression for the the kernel function $K(\bQ,\tau_z)$ 
introduced in Eq.~\eqref{eq.varphi} and used in the generalized \F\ matrix element in Eq.~\eqref{eq:keyexpression}.
\begin{widetext}
\begin{eqnarray} \label{eq:kernel}
&& K(\bQ,\tau_z) = \frac{1}{\gamma^- - \gamma^+} \frac{1}{Q^2} \nonumber \\ && \times 
\Bigg\{\left[(\alpha + \gamma^- \beta)e^{\Qin \tau_z} + (\beta + \gamma^-\alpha)e^{-\Qin 
\tau_z}\right]\times \nonumber \\ && \times \Bigg[ (\alpha + \gamma^+ \beta) [e^{(\Qin-iQ_z) 
\tau_z} - e^{-(\Qin-iQ_z)d}](\Qin+iQ_z) - (\beta + \gamma^+ \alpha) [e^{-(\Qin+iQ_z) 
\tau_z}-e^{(\Qin+iQ_z) d}](\Qin-iQ_z) \nonumber \\ && + \frac{\alpha + \gamma^+ \beta}{e^{-iQ_z 
c+\eta}-1} [1-e^{-(\Qin-iQ_z) d}](\Qin+iQ_z) -\frac{\beta + \gamma^+ \alpha}{e^{-iQ_z c+\eta}-1} 
[1-e^{(\Qin+iQ_z) d}](\Qin-iQ_z) \nonumber \\ && + \frac{1}{e^{-iQ_z c+\eta}-1}  [e^{(\Qin-iQ_z) 
D}-1](\Qin+iQ_z) -\frac{\gamma^+}{e^{-iQ_z c+\eta}-1}  [e^{-(\Qin+iQ_z) D}-1](\Qin-iQ_z) \Bigg] 
\nonumber \\ && + \left[(\alpha + \gamma^+ \beta)e^{\Qin \tau_z} + (\beta + \gamma^+\alpha)e^{-\Qin 
\tau_z}\right]\times \nonumber \\ && \times \Bigg[ (\alpha + \gamma^- \beta) [1 
-e^{(\Qin-iQ_z)\tau_z}](\Qin+iQ_z) - (\beta + \gamma^- \alpha) [1-e^{-(\Qin+iQ_z) 
\tau_z}](\Qin-iQ_z) \nonumber \\ && + \frac{\alpha + \gamma^- \beta}{e^{iQ_z c+\eta}-1} 
[1-e^{-(\Qin-iQ_z) d}](\Qin+iQ_z) - \frac{\beta + \gamma^- \alpha}{e^{iQ_z c+\eta}-1} 
[1-e^{(\Qin+iQ_z) d}](\Qin-iQ_z) \nonumber \\ && + [e^{(\Qin-iQ_z) D}-1](\Qin+iQ_z) - \gamma^- 
[e^{-(\Qin+iQ_z) D}-1](\Qin-iQ_z) \nonumber \\ && + \frac{1}{e^{iQ_z c+\eta}-1} [e^{(\Qin-iQ_z) D}- 
1](\Qin+iQ_z) - \frac{\gamma^-}{e^{iQ_z c+\eta}-1}   [e^{-(\Qin+iQ_z) D}-1](\Qin-iQ_z) \Bigg] 
\Bigg\}. \label{eq.fullkernel}
\end{eqnarray}
\end{widetext}
The definitions of the parameters $\a$, $\b$, $\gamma^\pm$, and $\eta$ are given in
Eqs.~\eqref{eq.defs1}-\eqref{eq.defs4}.

\begin{acknowledgments}
This research is supported by the Computational Materials Sciences Program funded by the U.S. 
Department of Energy, Office of Science, Basic Energy Sciences, under Award No. DE-SC0020129
(W.H.S., formalism, software development, \textit{ab initio} calculations, manuscript
preparation; F.G., project conception and supervision, formalism, manuscript preparation.)
The authors acknowledge the Texas Advanced Computing Center (TACC) at The University of Texas at 
Austin for providing HPC resources, including the Frontera and Lonestar5 systems, that have 
contributed to the research results reported within this paper. URL: http://www.tacc.utexas.edu. 
This research used resources of the National Energy Research Scientific Computing Center, 
a DOE Office of Science User Facility supported by the Office of Science of the U.S. Department of
Energy under Contract No. DE-AC02-05CH11231. W.H.S. was also supported by the Science and 
Technology Development Fund of Macau SAR (FDCT) (under grants No. 0102/2019/A2). 
W.H.S also acknowledges the Information and Communication Technology Office (ICTO) at 
the University of Macau and the LvLiang Cloud Computing Center of China for providing 
extra HPC resources to the code testing and benchmark, including the High Performance 
Computing Cluster (HPCC) and TianHe-2 systems.
\end{acknowledgments}

\bibliography{cite.bib}

\end{document}